\documentclass[12pt]{article}
\usepackage{amsfonts}
\usepackage{amsmath, amsthm}
\usepackage{epsfig}
\usepackage{algorithm}
\usepackage{algorithmic}
\usepackage{epic}
\usepackage{longtable}

\usepackage[a4paper]{geometry}
\usepackage{enumitem}
\usepackage{dsfont}
\usepackage{rotating}
\usepackage{mathtools}
\usepackage[table,xcdraw]{xcolor}
\usepackage{amsmath,verbatim,color,amssymb,epsfig}
\usepackage{bm}
\usepackage{amsfonts}
\usepackage{epsfig}
\usepackage{changepage}
\usepackage{multirow}
\usepackage{graphicx}
\usepackage{array}
\usepackage[table]{xcolor}
\usepackage{adjustbox,lipsum}
\definecolor{Gray}{gray}{0.80}
\usepackage{lscape}

\usepackage{rotating}
\usepackage{dsfont}
\usepackage[round,semicolon,authoryear]{natbib}
\usepackage[normalem]{ulem}
\DeclareMathOperator*{\argmax}{argmax}
\usepackage[colorlinks=true,urlcolor=blue,citecolor=purple,linkcolor=blue,bookmarks=true]{hyperref}
\usepackage{caption}

\def\eqx"#1"{{\label{#1}}}
\def\eqn"#1"{{\ref{#1}}}

\makeatletter 
\@addtoreset{equation}{section}
\makeatother  

\def\squarebox#1{\hbox to #1{\hfill\vbox to #1{\vfill}}}
\def\boxit#1{\vbox{\hrule\hbox{\vrule\kern6pt
          \vbox{\kern6pt#1\kern6pt}\kern6pt\vrule}\hrule}}

\newcolumntype{L}[1]{>{\raggedright\let\newline\\\arraybackslash\hspace{0pt}}m{#1}}
\newcolumntype{C}[1]{>{\centering\let\newline\\\arraybackslash\hspace{0pt}}m{#1}}
\newcolumntype{R}[1]{>{\raggedleft\let\newline\\\arraybackslash\hspace{0pt}}m{#1}}

\def\bzeta{\bm{\zeta}}

\usepackage{booktabs,multirow}
\usepackage{tikz}
\usetikzlibrary{shapes}

\usepackage{threeparttable}

\newcommand{\beginsupplement}{%
        \renewcommand{\thetable}{S\arabic{table}}%
        \renewcommand{\thefigure}{S\arabic{figure}}%
    \renewcommand{\thesection}{S\arabic{section}}%
    \renewcommand{\thesubsection}{S\arabic{section}.\arabic{subsection}}%
    }

\begin{document}

\baselineskip=20pt
\begin{center}
{\Large \bf ROMI:  A Randomized Two-Stage Basket Trial Design to Optimize Doses for Multiple Indications}
\end{center}
\begin{center}
{\bf Shuqi Wang$^{1}$, Peter F. Thall$^{1}$, Kentaro Takeda$^{2}$, and Ying Yuan$^{1,*}$}
\end{center}

\begin{center}
$^{1}$Department of Biostatistics, The University of Texas MD Anderson Cancer Center\\
Houston, TX, USA.\\
$^{2}$Astellas Pharma Global Development Inc., Northbrook, IL, USA.\\
{*Email: yyuan@mdanderson.org}\\
\end{center}

\begin{center}
\vskip 0.2 true cm

{\textbf{Abstract}}
\end{center}
\vskip 0.15 true cm

 
Optimizing doses for multiple indications is challenging. The pooled approach of finding a single optimal biological dose (OBD) for all indications ignores that dose-response or dose-toxicity curves may differ between indications, resulting in varying OBDs. Conversely, indication-specific dose optimization often requires a large sample size. To address this challenge, we propose a Randomized two-stage basket trial design that Optimizes doses in Multiple Indications (ROMI). In stage 1, for each indication, response and toxicity are evaluated for a high dose, which may be a previously obtained MTD, with a rule that stops accrual to indications where the high dose is unsafe or ineffective. Indications not terminated proceed to stage 2, where patients are randomized between the high dose and a specified lower dose. A latent-cluster Bayesian hierarchical model is employed to borrow information between indications, while considering the potential heterogeneity of OBD across indications. Indication-specific utilities are used to quantify response-toxicity trade-offs. At the end of stage 2, for each indication with at least one acceptable dose, the dose with highest posterior mean utility is selected as optimal. Two versions of ROMI are presented, one using only stage 2 data for dose optimization and the other optimizing doses using data from both stages. Simulations show that both versions have desirable operating characteristics compared to designs that either ignore indications or optimize dose independently for each indication.

\vskip 0.15 true cm
\noindent{KEY WORDS:} Bayesian hierarchical model; Dose optimization; Multiple indications; Project optimus; Randomization; Utility.

\baselineskip=24pt

\newpage

\section{Introduction} \label{sec:Introduction}

Conventional phase I oncology dose-finding designs originally were motivated by trials of cytotoxic agents, where the probabilities of toxicity, $\pi_T(d)$, and response, $\pi_R(d)$, increase with dose, $d$.  This may not hold for targeted molecules or immunotherapies,  where  $\pi_T(d)$ and $\pi_R(d)$
may take a variety of different shapes. For example, if the delivered dose is saturated in the patient,  the $\pi_R(d)$ curve initially increases with $d$ and then flattens to a plateau.
In such settings, a phase I maximum tolerated dose (MTD) is undesirable because lower doses achieve similar $\pi_R(d)$  but reduce $\pi_T(d)$  (\citealp{Sachs2016}). Thus,  conventional phase I  designs are unsuitable for most targeted agents \citep{Shah2021, CCR2023}.

To address these issues, the U.S. Food and Drug Administration (FDA) launched Project Optimus \citep{FDA2022}, and released guidance \citep{FDA2024-dose-optimization} to shift the dose-finding goal from identifying an MTD to determining an optimal biological dose (OBD) that maximizes a risk-benefit tradeoff. Following the FDA's recommendation to
 randomize patients among doses,
 several dose optimization designs including randomization recently have been proposed. 
 \citet{DROID2023} presented a design (DROID) combining the dose-ranging framework of non-oncology trials with oncology dose-finding designs. \citet{MERIT2024} developed a multiple-dose randomized trial (MERIT) design that optimizes dose based on toxicity, and provided an algorithm to determine sample size.
\cite{Genp12} proposed a generalized phase I-II design
that uses phase I-II criteria to identify a set of candidate doses based on response and toxicity,  randomizes patients among the candidates, and selects the best dose based on long-term treatment success. \cite{Genp123} extended that approach to a generalized phase I-II-III design, integrating it with a Phase III trial to further enhance the design's efficiency.  
 See \cite{Yuan2024} for a review.



 Identifying optimal doses for multiple indications is more difficult because one must account for the possibility that the indications may have different dose-outcome curves, and thus different OBDs. The FDA's guidance indicates that ``Different dosages may be needed in different disease settings or oncologic diseases based on potential differences in tumor biology, patient population, treatment setting, and concurrent therapies, among other factors" \citep{FDA2024-dose-optimization}. While a straightforward approach is to optimize dose independently for each indication,  this may lead to a very large sample size. 


This paper was motivated by an early phase trial at MD Anderson Cancer Center to identify OBDs of an anti-CD137 agonist in combination with pembrolizumab and nab-paclitaxel for treating metastatic solid tumors. Because the agonist induces responses in CD8+ T-cells, it was expected to complement and enhance the efficacy of the immune checkpoint blockade pembrolizumab.  Doses of pembrolizumab and nab-paclitaxel were fixed at 200 mg and 220 mg/m$^2$, respectively. The MTD of the CD137 agonist was established in an all-comer dose escalation trial with several indications. The investigator was interested in conducting a dose optimization trial by randomizing patients between the MTD and a lower dose. Four indications were studied: esophageal and gastric cancer, head and neck cancer, Her2-negative breast cancer, and ovarian cancer. Since the treatment might be ineffective in some indications, one aim was to minimize the sample sizes of indications with poor results.

To efficiently identify an OBD for each indication in this setting, the two-stage basket trial design, ROMI, described in this paper was developed. 
Denote indications by $I_1,\cdots,I_K$ and index stages by $s=1,2.$ We consider settings where an MTD of a new agent has been provided, possibly based on an earlier phase I trial in one $I_k$ or all-comers. The goal is to identify an OBD for each $I_k$ based on binary toxicity and response.
Stage 1 of ROMI  focuses on screening a high dose, $d_H$, which is the MTD that has been provided, in each $I_k$.
Accrual to an $I_k$ is terminated if it is found that $\pi_R(d_H)$  is unacceptably low or $\pi_T(d_H)$
 is unacceptably high, compared to fixed limits specified for  $I_k$.
In stage 2,  the goal is to select an OBD for each $I_k$, with
patients randomized between $d_H$ and a prespecified lower dose, $d_L$, while doing safety and futility monitoring for each dose in each $I_k$. To select OBDs, a ROMI design requires elicited numerical utilities of the four possible (toxicity, response) outcome pairs to compute a decision criterion. A Bayesian hierarchical model is assumed that allows the $I_k$'s to have different OBDs, and borrows information between the $I_k$'s. For each $I_k$, the OBD is the acceptable dose with maximum posterior mean utility. We present two versions of the ROMI design. The first version uses only the randomized stage 2 data to select OBDs.   The second version uses the data from both stages, based on an extended hierarchical model accounting for possible bias due to drift of $d_H$ effects between stages 1 and 2.

In Section \ref{sec:Method1}, we present the first version of the ROMI design, including the hierarchical model, descriptions of each stage,  an illustrative example, and guidelines for determining sample size.
Section \ref{sec:Method2} presents the second version of the ROMI design, including a model elaboration to account for possible drift of $d_H$ effects between stages. Section \ref{sec:simulation} reports simulations that evaluate the operating characteristics of the ROMI designs and compare them to designs that choose one dose for all $I_k$'s or conduct separate trials within the $I_k$'s.
We close with a discussion in Section \ref{sec:Discussion}.

\section{Notation and Design Elements} \label{sec:Method1}    
While 
a  ROMI design can accommodate more than two doses, for simplicity and to control overall sample size, we will restrict attention to the case of two doses,  $\{d_L, d_H\}$. A ROMI design with more than two doses is described in Supplementary Material S1. 
We consider settings where dose evaluation is based on binary toxicity, $Y_T$, and binary response, $Y_R$.   
In stage 1,  all patients are treated with $d_H$, and  $I_k$'s for which  $d_H$ is unsafe or ineffective are screened out. $I_k$'s passing stage 1 screening go to stage 2, where patients are randomized between $d_H$  and $d_L$, and each dose is screened in each $I_k$.  At the end of stage 2, for each $I_k$ with at least one acceptable dose, the OBD  is defined as the dose maximizing posterior mean utility.

The remainder of this section will describe the first version of ROMI, where only stage 2 data are used to choose  OBDs.  The second version, which uses both the stage 1 and stage 2 data to choose OBDs, is presented in Section 3.

\subsection{Stage 1 Dose Screening} {\label{subsec:stage1}}

Denote the maximum stage $s$ sample size for dose $d_{\ell}$ in $I_k$  
 by $N_{\ell,k, s}$.  Because only $d_H$ is evaluated in stage 1,  $N_{L,k,1}=0$ for all $k$.
For   $I_k,$  when the maximum sample size $N_{H,k,1}$ of $d_H$  in stage 1 is reached, the acceptability of $d_H$
is evaluated using two screening rules,  constructed
using the   approach of \cite{Thall1998} and \cite{Zhou2017}, which is used by numerous designs.
Let $X_{T, H, k, 1}$ denote the number of toxicities and $X_{R, H, k, 1}$  the number of responses among the $N_{H, k, 1}$ patients with indication $I_k$ in stage 1.  Denote
the stage 1 count data by
$\mathcal{D}_1=\{(N_{H, k,1}, X_{T, H, k, 1},  X_{R, H, k, 1}), k=1,\dots,K\},$ and
the marginal outcome probabilities  $\pi_{j, \ell, k} = \Pr(Y_j =1  \mid d_{\ell}, I_k)$ for $j=R,T$,   $\ell$ = $H, L$, and $k=1,\cdots,K$.  For each $I_k,$
$\overline{\pi}_{T,k}$ denotes a fixed maximum acceptable toxicity probability, and   $\underline{\pi}_{R,k}$  a fixed minimum response probability, elicited from the clinical investigators.  The  values
of $\overline{\pi}_{T,k}$ may be the same or similar across indications, but values of $\underline{\pi}_{R,k}$ may vary substantially with $k$ due to qualitatively different definitions of response
and therapeutic expectations across the $I_k$'s.  Accrual to
   $I_k$ is terminated at the end of stage 1  if $d_H$  is found  likely to be excessively toxic, using the posterior  safety criterion
    \begin{equation}
        \text{Pr}(\pi_{T,H,k} > \overline{\pi}_{T,k}\mid \mathcal{D}_1) > c_{T,k,1},     \label{eqn: toxicity monitoring}
    \end{equation}
or if it is found likely to be inefficacious, using the posterior futility criterion
       \begin{equation}
        \text{Pr}(\pi_{R,H,k} <  \underline{\pi}_{R,k} \mid \mathcal{D}_1) > c_{R,k,1}.  \label{eqn: futility monitoring}
    \end{equation}
The cutoffs $c_{T,k,1}$ and $c_{R,k,1}$ are fixed at values such as 0.90 or 0.95, calibrated by preliminary simulations to obtain good operating characteristics, 
including a high probability of stopping accrual to indications where $d_H$ is too toxic, with $\pi_{T,H,k}^{true} > \overline{\pi}_{T,k}$, or inefficacious, with  $\pi_{R,H,k}^{true} < \underline{\pi}_{R,k}$.
 
 To evaluate  posterior probabilities in the stage 1 monitoring rules (\ref{eqn: toxicity monitoring}) and
 (\ref{eqn: futility monitoring}), we assume  beta-binomial models,
 with  non-informative priors $\pi_{j,H, k} \sim Beta(0.1, 0.1)$,  
 and likelihoods
$$
X_{j,H, k, 1} \mid \pi_{j,H,k} \sim Binom(N_{H,k, 1},\ \pi_{j,H,k}), \quad j=R,T.
$$
By conjugacy, the posteriors are
 $$
 \pi_{j, H, k}\mid \mathcal{D}_1 \sim  Beta(0.1 + X_{j,H,k,1},\ 0.1 + N_{H,k, 1} -X_{j,H,k,1}).
 $$
The monitoring rules also may be applied before the end of stage 1,  e.g., after evaluating $N_{H,k, 1}/2$ patients in $I_k,$ and at $N_{H,k, 1}.$
Each $I_k$  with acceptable response and toxicity rates for $d_H$ at the end of  stage 1 is moved to stage 2, otherwise
no dose is chosen for $I_k$.

\subsection{Stage 2 Dose Optimization} {\label{subsec:stage2}}
In stage 2,  patients are randomized between $d_H$ and $d_L$. The aim is to identify an OBD for each $I_k$,
 based on indication-specific  utilities
$U_k(y_T,y_R)$  for $y_T, y_R \in \{0,1\}$ and $k=1,\cdots,K.$
For each $I_k$, one may  establish $U_k(y_T,y_R)$   by  setting $U_k(0,1)$ = 100  for the best  outcome (no toxicity, response),
$U_k(1,0)$ = 0 
for the worst outcome (toxicity, no response),  and eliciting $U_k(0,0)$ and $U_k(1,1)$ from the physicians.
Table \ref{tab:utility-example} gives a numerical example of utilities for four indications.
Utility-based phase I-II designs are given by \cite{Thall2012}, \cite{Guo2017},   and \cite{UBOIN2019}, among many others.

To do utility-based dose optimization for each $I_k$ based on the randomized stage 2 data, denote the joint elementary outcome probabilities  for dose $d_{\ell}$ in $I_k$ by
\begin{equation}
p_{\ell,k}(y_T,y_R)
=\Pr(Y_T = y_T, Y_R=y_R \mid d_{\ell}, I_k), \ \  {\rm for}\ \ \ y_T, y_R\in\{0,1\}.
\label{joint}
\end{equation}
The mean utility of $d_{\ell}$ in  $I_k$ is the probability weighted average
\begin{equation} \label{eqn:utility}
\overline{U}_{\ell, k} = \sum_{y_T=0}^1 \, \sum_{y_R=0}^1 U_k(y_T,y_R)\,p_{\ell,k}(y_T,y_R).
\end{equation}
Following the utility-based BOIN12 design \citep{BOIN122020}, we take a quasi-binominal likelihood approach by defining  standardized mean utilities $Q_{\ell, k}=\overline{U}_{\ell,k}/100$, called  \lq\lq quasi-probabilities" because
they take values between 0 and 1.
For each $d_{\ell}$ an d $I_k,$ let $X_{\ell,k}(y_T,y_R)$ denote the number of patients in stage 2 who experience the joint outcome $(y_T,y_R)$,  and denote the vector of  counts for the four elementary outcomes by
\begin{equation}\label{dataX}
\boldsymbol{X}_{\ell,k} = (\,X_{\ell,k}(0,1),X_{\ell,k}(0,0),X_{\ell,k}(1,1),X_{\ell,k}(1,0)\,),
\end{equation}
 with corresponding  joint probability vector $\boldsymbol{p}_{\ell,k}$.
 Thus,  $\boldsymbol{X_{\ell,k}} \sim Multinomial(N_{\ell,k,2}, \boldsymbol{p}_{\ell,k})$ for each $d_{\ell}$
    and $I_k.$   Given the stage  2 data,  we define  normed utility-weighted  average counts
\begin{equation} \label{eqn:standersized_utility_mean}
    Z_{\ell, k} = \frac{1}{100}\sum_{y_T=0}^1\sum_{y_R=0}^1 U_{k}(y_T,y_R)X_{\ell, k}(y_T,y_R).   \nonumber
\end{equation}
Each $Z_{\ell, k}$ has domain  (0,\ $N_{\ell,k,2}$), and may take non-integer values.
It
may be interpreted as the number of \lq\lq quasi-events\rq\rq\ among the $N_{\ell, k,2}$ patients with indication $I_k$ treated with  $d_{\ell}$ in stage 2.
Given the quasi-probability $ Q_{\ell, k}$,
we denote the distribution of  $Z_{\ell, k}$  induced by the multinomial distribution of $\boldsymbol{X_{\ell,k}}$
 by
 $
 Z_{\ell, k} \sim  {\rm Quasi}-{\rm Binom} (N_{\ell, k, 2}, Q_{\ell, k}).$

To use the stage 2 data to select OBDs, we proceed as follows.  We accommodate heterogeneity among indications and facilitate borrowing information between indications by  introducing a vector of {\it latent cluster variables} $\bzeta$ = $(\zeta_1,\cdots,\zeta_K)$ \citep{Blast2018, Chen_Lee2019, Takeda2022}, 
 where  $\zeta_k$ =  $I[Q_{H,k} \leq Q_{L,k}]$,    the indicator that $d_L$ has higher mean utility  than $d_H$ in  $I_k.$
Let $N(\mu,\sigma^2)$ denote a normal distribution with mean $\mu$
and variance $\sigma^2$, and  $IG(a,b)$ an  inverse gamma distribution with
parameters $a$ and $b$. Recall that $Z_{\ell,k}$ is the number of quasi-events and $Q_{\ell,k}$ is the
quasi-probability for $I_k$ and  $d_{\ell}$ in stage 2.
Denote $\theta_{k}$ = 
${\rm logit}(Q_{L,k})$  -  
${\rm logit}(Q_{H,k}),$
 the $d_L$-versus-$d_H$ effect in $I_k$, where  ${\rm logit}(q) = {\rm log}\{q/(1-q)\}$ for $q\in[0,\ 1]$.  Thus, $\theta_{k}$ is  a function of $\overline{U}_{L, k}$, $\overline{U}_{H, k}$,
and  the probability vectors $\{p_{\ell, k}\}$.
 For the  stage 2 data,
we assume  the  Bayesian hierarchical model 
\begin{eqnarray} \label{eqn:BHM-v1}
    \begin{aligned}
     &Z_{\ell, k} \mid Q_{\ell, k} \sim Quasi{\rm -}Binom(N_{\ell, k, 2}, Q_{\ell, k}),\ \ {\rm for}\ \  \ell = L, H, \ k=1,\cdots,K, \\
    & \theta_k \mid \zeta_k=g \sim iid \  N(\mu_g,  \tau^2), \ \ \ {\rm for}\ \ g =0,1, \ \  {\rm and\ each} \ k=1,\cdots,K.
    \end{aligned}
\end{eqnarray}
  For priors, we assume 
\[
\mu_g  \sim N(\widetilde{\mu}_g, \widetilde{\tau}_g^2),  \ {\rm for}\ \ g =0,1, \ {\rm and}\ \  \tau^2  \sim IG(a,b),
\]
\[
Q_{H,k} \sim  Beta(c,d),  \ \ \zeta_k \sim Bernoulli(q),  \ \ {\rm and}\ \ q \sim Beta(e,f),
\]
with $\widetilde{\mu}_g, \widetilde{\tau}_g^2, a, b, c, d, e, f$ as fixed hyperparameters.
Since 
$\boldsymbol{p}_{L,k}$ and
$\boldsymbol{p}_{H,k}$
 contribute to the stage 2 likelihood only through the quasi-probabilities $Q_{L,k}$ and $Q_{H,k}$, one only needs to specify priors on these
to complete the model. Since  
normal priors are specified on $\theta_k$  for each $k$, 
the model is completed by specifying priors on the $Q_{H,k}$'s.

Hyperparameters may be established by applying the approach of \cite{Thall2012} and \cite{Guo2017}. 
To do this, expected response and toxicity probabilities are elicited from the clinicians for each combination  $(d_{\ell},I_k)$.
 These provide a basis for calculating a range of utility differences between $d_{L}$ and $d_H$ on the logit scale, i.e., for the $\theta_k$'s. One may set $\widetilde{\mu}_0$ to the mean in the subset where $\theta_k < 0$, and set $\widetilde{\mu}_1$  to the mean in the subset where $\theta_k \geq 0$. Once $\widetilde{\mu}_0$ and  $\widetilde{\mu}_1$ are established, one may assume a coefficient of variation of 2, which sets $\widetilde{\tau}_g = 2\mu_g$ \citep{Guo2017}.  
The shrinkage parameter $\tau^2$ can be assigned an inverse gamma prior, such as, IG($0.0001,$ $0.0001$).  \citet{Gelman2006} and \citet{CBHM2018} noted that the IG($\epsilon$,$\epsilon$) with $\epsilon \rightarrow 0$ does not represent a non-informative prior, but instead imposes strong shrinkage when the number of elements in the hierarchy (indications in our context) is small (e.g., $\le 6$) unless the heterogeneity between indications is extremely large.  Under our model, this potential problem is mitigated by using $\bzeta$ to partition the indications into ${\cal I}^0$ and ${\cal I}^1$. Since indications in each of these subsets are likely to be homogeneous, the strong shrinkage effect of the prior often enhances the model's performance. As a sensitivity analysis, we consider a Half-Cauchy distribution prior for $\tau^2$ in Section \ref{subsec:sensitivity}.

For each $I_k$ where $d_H$ passes the stage 1 screening,  in stage 2  patients are randomized between $d_H$ and $d_L$. If $R$ interim screening analyses are carried out for $I_k$ in stage 2,  let $n_{\ell, k, 2, r}$ denote the interim sample size for the $k^{th}$ indication at the $r^{th}$ stage 2 look.  Let $\mathcal{D}_{2,r}$ denote the data at $r^{th}$ interim look, and  $\mathcal{D}_2$  the final data from stage 2. At the $r^{th}$ interim analysis,  $(Y_T,Y_R)$  are evaluated for all patients treated at each dose, and a dose is terminated if it is excessively toxic per criteria (\ref{eqn: toxicity monitoring}) or ineffective per criteria (\ref{eqn: futility monitoring}). To reduce bias, futility monitoring relies solely on the stage 2 data. In contrast, safety monitoring pools the stage 1 and stage 2 data,   assuming toxicity probabilities will not change between stages.

At the end of stage 2, for each  $I_k,$ when the maximum stage 2 sample sizes $N_{L,k,2}$ and $N_{H,k,2}$ are reached for the two doses,
 a final analysis is conducted to determine the OBD.  The toxicity monitoring  rule (\ref{eqn: toxicity monitoring}) is applied for each dose based on $\mathcal{D}_1 \cup \mathcal{D}_2$, and futility monitoring   is done based on the stage 2 data  using the rule
$\text{Pr}(\pi_{R,\ell, k} < \underline{\pi}_{R,k}|\mathcal{D}_2) > c_{R,k,2}.$
 For  $I_k$, the OBD is the dose that passes both the toxicity and response requirements and maximizes the posterior mean standardized utility, estimated under the Bayesian hierarchical model.  The dose optimization criterion in $I_k$ is  denoted by
    \begin{equation}\label{eqn:dose-recommend}
       \text{OBD}_k = \argmax_{\ell = L,H}\ \widehat{Q}_{\ell,k} = \argmax_{\ell = L,H}\ E\{ {Q}_{\ell,k}  \mid {\cal D}_2\}.
        \end{equation}
\subsection{ Graphical Illustration of Trial Conduct}
\begin{figure}[H]
    \centering
    \includegraphics[width=0.97\textwidth]{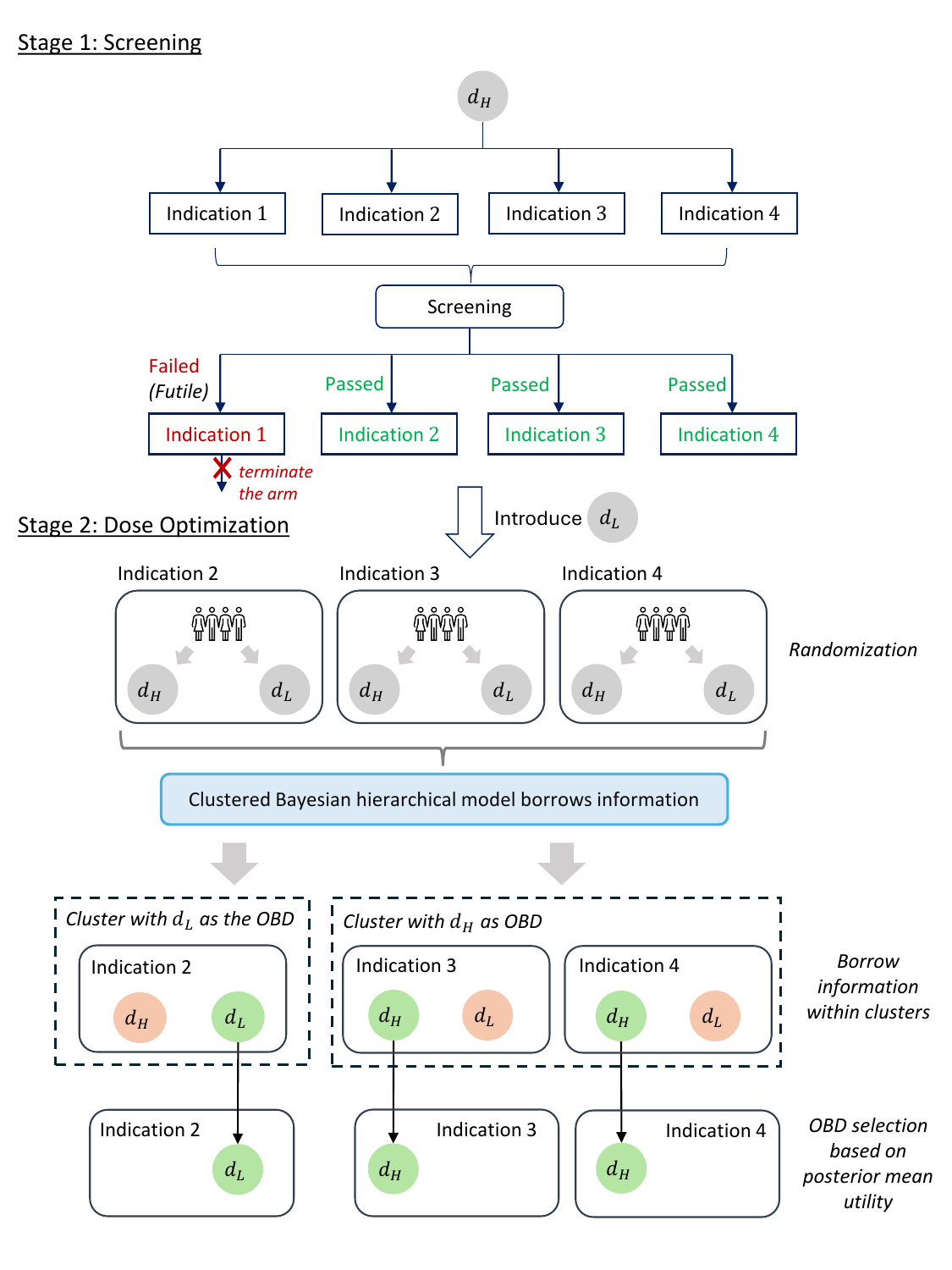}
     \vspace*{-7mm}
    \caption{A ROMI design example with four indications and two doses, $d_H$ and $d_L$. OBDs are indicated by green circles. 
}
    \label{fig:ROMI_example}
\end{figure}

{\label{subsec:example}}
Figure \ref{fig:ROMI_example} presents a schematic of trial conduct using the ROMI design to determine the OBD, if it exists, between two doses $d_L$ and $d_H$ for each of four disease subtypes (indications).
In stage 1, all patients are treated with $d_H$, and toxicity and response are monitored for each $I_k$. Due to an unacceptably low response rate with $d_H$, $I_1$ is dropped, while  $I_2,$ $I_3,$ and $I_4$ are moved forward to stage 2, where
 patients are randomized between $d_H$ and $d_L$.
A final analysis is conducted to evaluate each dose's safety, response rate, and mean utility. For $I_2$, both doses have acceptable toxicity and response rates, with  $d_L$  selected as the OBD based on posterior mean utility. For $I_3$ and $I_4$, $d_H$ is selected as the OBD due to its higher posterior mean utility.
Thus,  the ROMI design does not identify an OBD for $I_1,$ identifies $d_L$  as the OBD for $I_2$, and identifies $d_H$ as the OBD for $I_3$ and $I_4$.

\subsection{Sample Size Determination} 

The sample size for each $I_k$ in stage 1 of a ROMI design may be determined to control the false negative decision probability of the futility stopping rule (\ref{eqn: futility monitoring}).
To do this, suppose that, for each $I_k$, a desirably high response probability $\underline{\pi}_{R,k}+\delta_{R,k}$ can be specified, say for  $\delta_{R,k}$ =  0.15,  0.20 or  0.25. The cut-off $c_{R,k,1}$ and sample size $N_{H,k,1}$ may be calibrated together by simulation so that, for true response probability $\pi_{R,k}^{true} = \underline{\pi}_{R,k}+\delta_{R,k}$, the false negative early stopping probability is no larger than a specified small value, such as  0.10 or  0.05. In practice, one may fix  $c_{R,k,1}$ at a large value, such as  0.90 or 0.95, and do a monotone search for the smallest $N_{H,k,1}$ that ensures the specified false negative early stopping probability.

To determine the sample size for each indication in stage 2, one can first apply the MERIT design \citep{MERIT2024}, which gives a structured approach for calculating sample size in randomized phase II  dose optimization studies. To do this, for each $I_k$, one may begin  by  specifying the lower limit $\underline{\pi}_{R,k}$, a desirably high response probability $\underline{\pi}_{R,k}+\delta_{R,k}$
with $\delta_{R,k}$ =  0.15,  0.20, or  0.25 as above,
  an upper  toxicity probability limit $\overline{\pi}_{T,k}$, and a desirably low toxicity probability $\overline{\pi}_{T,k}-\delta_{T,k}$.
One then specifies a maximum level, such as 0.10 or 0.15,  for the probability of incorrectly accepting an undesirable dose (type I error rate), and a minimum level, such as 0.60, 0.70, or 0.80,
for the probability of correctly choosing an acceptable dose (power). The MERIT sample size $N_{\ell,k,2}^M$  for dose $d_{\ell}$ and indication $I_k$  may be determined by a numerical search, to find the smallest value that controls the type I error while achieving the desired power. Since MERIT assumes equal randomization, for a ROMI design, one may restrict the randomization by requiring $N_{H,k,2}^M = N_{L,k,2}^M$.  Software for calculating sample size using the MERIT design is available at \citet{trialdesign}.

The MERIT design method may be used to determine the sample size for each indication independently. Compared to a randomized trial assuming homogeneity,
however, the ROMI design allows information borrowing between indications, which may reduce the planned overall sample size while still preserving a given level of accuracy in selecting the OBDs at the end of the trial.   To exploit this, the stage 2 sample sizes $\{N_{L,k,2}^M\}$  and $\{N_{H,k,2}^M\}$  obtained from the MERIT design may be adjusted by simulating the ROMI design to achieve the desired level of reliability in the final dose selections. For example, with $K=2$ indications and initial stage 2 sample sizes $(N_{\ell,1,2}^M, N_{\ell,2,2}^M)=(30, 25)$, simulations of the trial using the ROMI design can be conducted with specified stage 1 sample sizes $\{N_{H,k,1}\}$, determined as described above,  and several combinations of stage 2 sample sizes, for example,  $(N_{\ell,1,2},N_{\ell,2,2})=(30, 25), (25,25), (20,25),(30,20),(25,20),(20,20)$, to assess operating characteristics. The sample size chosen for stage 2 is based on the tradeoff between the accuracy of the final OBD selection for each $I_k$ and total trial sample size $N=\sum_{k=1}^K(N_{H,k,1}+N_{H,k,2}+N_{L,k,2}).$
If desired, the $\{N_{H,k,1}\}$ values may be adjusted and the trial simulations repeated.

\section{Using Data from Both Stages} \label{sec:Method2}
Combining data on $d_H$ from both stage 1 and stage 2  may improve the estimate of $d_H$-versus-$d_L$ effects for the OBD selection in each indication.
This is straightforward when it is reasonable to assume that the data from stages 1 and 2 are exchangeable: simply pool the data from both stages when calculating  $\boldsymbol{X}_{\ell,k}$ in (\ref{dataX}). However, since there is no randomization in stage 1, and patients are randomized to $d_H$ or $d_L$ in stage 2, there might be  drift in the effect of $d_H$ on the outcomes between stages,  possibly due to temporal changes in patient characteristics or unknown factors. In this case, simply pooling the data results in bias.

To include stage 1 data on $d_H$ and account for potential temporal drift, we extend the Bayesian hierarchical model, referred to as Version 2 of ROMI. The joint distributions $p_{H,k}(y_R,y_T)$,  defined earlier, are elaborated to be stage-specific distributions  $p_{H,k,s}(y_R,y_T)$  for $s=1$ and 2 and all $I_k.$  This  produces stage-specific mean utilities $\overline{U}_{H,k,1}$
   and $\overline{U}_{H,k,2}$,
    quasi-probabilities $Q_{H,k,1}$ and $Q_{H,k,2}$, and between-dose effects
       $\theta_{k,s}$  = ${\rm logit}(Q_{L,k,s})  -  {\rm logit}(Q_{H,k,s})$.
 Since no patients are treated with $d_L$ in stage 1, however,
for the stage 2 selection only $\theta_{k,2}$ is relevant for each $I_k.$
We account for the stage by letting $Z_{\ell,k,s}$  denote the number of quasi-events and $Q_{\ell,k,s}$  the standardized utility for $I_k$ at dose $d_{\ell}$ in stage $s$ = 1 or 2.   Because $d_L$ is not evaluated in stage 1, each  $Z_{L,k,1} = 0$.  Thus, for $d_L,$ only $Z_{L,1,2}, \cdots, Z_{L,K,2}$ are defined and used in  the stage 2 decisions.

To model stage 1 data on $d_H$ and stage 2 data on $\{d_L,d_H\}$, we assume  an
extended Bayesian hierarchical model that accounts for the use of stage 1 quasi-event values $Z_{H,k,1}$ with the stage 2 values $Z_{L,k,2}$ and $Z_{H,k,2}$. For each $I_k,$ denoting the drift parameter
$\beta_k$ = ${\rm logit}(Q_{H, k, 1}) - {\rm logit}(Q_{H, k, 2}),$  
we assume
\begin{eqnarray} \label{eqn:BHM-v2}
    \begin{aligned}
     &Z_{H, k, 1} \mid Q_{H, k, 1} \sim Quasi{\rm -}Binom(N_{H, k, 1}, Q_{H, k, 1}), \quad \quad\quad\quad\quad\quad\quad (\rm \ {\bf Stage\ 1}\ )\ \\
     & Z_{\ell, k, 2} \mid Q_{\ell, k, 2} \sim Quasi{\rm -}Binom(N_{\ell, k, 2}, Q_{\ell, k, 2}), \ \ {\rm for}\ \ \ell = L,H, \quad\ \ (\rm \ {\bf Stage\ 2}\ )\\
     &\theta_{k,2}  = {\rm logit}(Q_{L,k,2})  -  {\rm logit}(Q_{H,k,2}), \\
       & \theta_{k,2} \mid \zeta_k=g \sim iid\ N(\mu_g,  \tau^2), \ \ \ {\rm for}\ \ g =0,1, 
    \end{aligned}
\end{eqnarray}
with  priors 
\begin{eqnarray} \label{eqn:BHM-v2-prior} \nonumber
    \begin{aligned}
    & \beta_k \sim\   \omega N(0,\sigma^2_{spike}) + (1-\omega)N(0,\sigma^2_{slab}), \\
    & \mu_g  \sim N(\widetilde{\mu}_g, \widetilde{\tau}_g^2), \ \ {\rm for}\ \ g =0,1, \ \ \tau^2  \sim IG(a,b), \ \  {\rm and} \ \ \omega \sim U[0,\ 1],\\
        & Q_{H,k,2}  \sim  \ Beta(c,d), \ \ \zeta_k \sim  \ Bernoulli(q), \ \ {\rm and}\ \  q \sim Beta(e,f). \\
    \end{aligned}
\end{eqnarray}
The variance $\sigma^2_{spike}$  should be set to a small value, such as 0.01, to concentrate the prior spike's mass near 0, while $\sigma^2_{slab}$  should be much larger than $\sigma^2_{spike}$ to allow a broader range of non-zero values for $\beta_k$. Following \citet{Gelman08} and \citet{Guo2017},  we regularize the prior so that the typical variation of an input variable is unlikely to cause a dramatic change in the response variable. E.g., $\beta_k=1$ corresponds to between-stage drift in $Q_{\ell, k}$  from  0.30 to 0.54. Based on the utility of $I_1$  in Table \ref{tab:utility-example}, a change of 0.24 in $Q_{\ell, k}$ corresponds to large shifts of 0.6 in  $\pi_{T, \ell, k}$
or of 0.4 in $\pi_{R, \ell, k}$.   Since it is very unlikely that between-stage drift would induce such large changes in the $\pi_{j, \ell, k}$'s, 
 we set $\sigma^2_{slab}=0.5^2$ to ensure that a change in $\beta_k$ from one standard deviation (sd) below to one sd above the mean is unlikely to cause a change of $Q_{\ell, k}$ exceeding 0.24.

Decision rules for version 2 of ROMI are as in Section \ref{subsec:stage2}. The only difference is that the posterior mean of the standardized utility is estimated under the extended model (\ref{eqn:BHM-v2}), using data from both stages and accounting for possible drift of $d_H$ effects between stages.

\section{Simulation Studies}  \label{sec:simulation}
\subsection{Simulation settings}
This section reports simulations to evaluate operating characteristics of the ROMI designs, and designs that either ignore the $I_k$'s or conduct separate trials within $I_k$'s.  We consider settings with $K=4$,  using dose  acceptability limits  $\overline{\pi}_{T,k}=0.40$ and $\underline{\pi}_{R,k}=0.25$ for all  $k$.
For each $I_k$, the maximum stage 1 sample size is 14, and the maximum stage 2 sample size per dose is 20, with one interim analysis performed when the sample size for each dose reaches 10. We constructed scenarios by varying the number of effective $I_k$'s and the OBD for each $I_k$. The utility table used for all $I_k$'s corresponds to that given for $I_1$ in Table \ref{tab:utility-example}. 
To characterize association between $Y_R$ and $Y_T$, for each dose $\ell$ = $L, H$ and $I_k$, given  marginal probabilities
$\pi_{T,\ell,k}$ and $\pi_{R,\ell,k}$, we solved for the joint probabilities 
$\{p_{\ell,k}(y_T,y_R)\}$ so that 
\[
\phi = \frac{p_{\ell,k}(0,0)p_{\ell,k}(1,1) - p_{\ell,k}(1,0)p_{\ell,k}(0,1)}{\{\pi_{R,\ell,k}(1-\pi_{R,\ell,k})\pi_{T,\ell,k}(1-\pi_{T,\ell,k})\}^{1/2} }  \ = \ .25.
\]
 We set $\widetilde{\mu}_0=-0.05$, $\widetilde{\mu}_1=0.05$, and $\widetilde{\tau}_0 = \widetilde{\tau}_1 = c = d = e = f =0.1$, $\tau^2 \sim IG(0.0001, 0.0001)$,  $\sigma^2_{spike} = 0.01$, and $\sigma^2_{slab} = 0.5^2$.

We denote the first version of ROMI design, which uses only stage 2 data for dose optimization, by ROMI-v1, and the second version, which uses data from both stages to optimize dose, by ROMI-v2. To assess the impact of clustering $I_k$'s showing similar dose-outcome probabilities, we define the ROMI-v1-NC design to have the same structure as ROMI-v1 but using the Bayesian hierarchical model without clustering. The first comparator is the Pool design, which ignores  $I_k$'s and determines the same OBD for all $I_k$'s based on the utility under a beta-binominal model,  $Z_{\ell} \sim Binom(\sum_k n_{\ell, k}, Q_{\ell})$ with a conjugate prior $Q_{\ell} \sim Beta(0.1,0.1)$. The second comparator is the Independent design,  a two-dose randomized design done independently for each $I_k$, with the utility of each arm modeled using a beta-binominal model, $Z_{\ell, k} \sim Binom(n_{\ell, k}, Q_{\ell, k})$ with a conjugate prior $Q_{\ell, k} \sim Beta(0.1,0.1)$.

For a fair comparison, the total maximum sample size for all designs was set to $N=216$. In the Independent design, patients within each  $\{I_1, I_2, I_3, I_4\}$
were randomized between the two doses, with a maximum of 27 patients per dose. For each $I_k$, one interim analysis was conducted after 14 patients. For the Pool design, one interim analysis was conducted when 108 patients were evaluated. The same interim stopping rules were used for all designs, with cutoffs set to $c_{T,k,1}=c_{R,k,1}=c_{R,k,2}=0.95$. A total of 2000 simulations were conducted for each combination of design and scenario.

\subsection{Simulation results}\label{subsec:simu_results}
Table \ref{tab:simulation-results} summarizes simulation results of the Pool, Independent, ROMI-v1-NC, ROMI-v1, and ROMI-v2 designs across 11 scenarios, assuming no drift in the effect of $d_H$ between stages. In Scenario 1, where no doses are effective for any $I_k$, the Pool design correctly stops all trials with no OBD selected for any $I_k$ 100\% of the time.  For each $I_k$, the stopping percentage  with no dose selected is $100$ $-$ ( \% select $d_H$ + \% select $d_L$). The stopping percentage is about 94\% for the Independent design and  98\% for designs using the ROMI framework, including ROMI-v1-NC, ROMI-v1, and ROMI-v2. Compared to the Pool and Independent designs, the ROMI designs provide substantial sample size savings, with about 42  fewer subjects than the Pool design and 56  fewer than the Independent design. This large sample size reduction for the ROMI designs in Scenario 1, where neither dose is effective, is due to the interim screening rule for $d_H$ applied by the ROMI designs after stage 1.

In scenarios 2 and 3, only $I_1$ responds to treatment. In Scenario 2, $d_H$ is the true OBD for $I_1$. ROMI-v2 and Independent design have the highest OBD correct selection percentages (CSPs), 69.7\% and 69.8\%, respectively. The ROMI-v1 and ROMI-v1-NC designs have CSPs 4.7\% and 5.2\% lower than ROMI-v2. The Pool design, which ignores indications, stopped 93.7\% of trials with a CSP of just 5.5\%. Compared to the Independent design, the ROMI designs save 42 subjects on average. A similar sample size saving is seen in scenario 3, where the true OBD for $I_1$ is $d_L$. In this case, the Pool design has a very low CSP of 4.3\%, while the Independent and ROMI designs have similar CSPs of around 68\%.

In scenarios 4 and 5, two indications respond to the treatment. In scenario 4, where the true OBD is $d_H$ for $I_1$ and $I_4$, the ROMI designs outperform the Pool and Independent designs in both CSP and sample size saving. ROMI-v2 has a CSP of 70.7\%, comparable to the highest CSP of 71.6\% achieved by ROMI-v1-NC. It is about 2\% higher than ROMI-v1 and Independent, and 7.7\% higher than Pool. The ROMI designs save an average of 27 subjects compared to the Independent design and 42 subjects compared to the Pool design. In scenario 5, where $d_L$ is the true OBD, ROMI designs save 28 subjects compared to the Independent design and up to 59 compared to the Pool design. The CSP of ROMI-v2 is 69.2\%, and ROMI-v1 is 68.3\%, similar to Independent but about 2.5\% lower than ROMI-v1-NC and Pool, both with a CSP of 71.7\%. Since the Pool design ignores indications and selects the same OBD for all $I_k$, it has high false positive rates of selecting ineffective doses for non-responsive indications. In scenario 4, Pool selects an ineffective dose for $I_2$ and $I_3$ 87.4\% of the time, rising to 96.1\% in scenario 5. In contrast, the probability of selecting an ineffective dose for these indications is 5.6\% with Independent and 1.8\% with ROMI designs.

Across scenarios 6, 7, and 8, where $I_1$ is insensitive to treatment, the ROMI designs save an average of 14 subjects compared to the Independent design and 35 compared to the Pool design. In scenario 6, $d_H$ is the true OBD for $I_2,$ $I_3,$ and $I_4$, while in scenario 7, $d_L$ is the OBD. Under heterogeneous scenarios where some indications are non-responsive and responsive $I_k$'s share the same OBD, ROMI-v1-NC and ROMI-v2 show larger CSPs compared to the Independent design, with increases of 5\% and 2.5\% in scenario 6, and 6\% and 2.8\% in scenario 7. These improvements show the benefit of borrowing information across indications. ROMI-v1 has a CSP similar to the independent design. The Pool design is effective in selecting the OBD for responsive indications but fails to terminate ineffective doses for non-responsive $I_1$, with a 100\% chance of choosing an ineffective dose.
ROMI-v1-NC has the highest CSP due to the strong shrinkage but underperforms in scenarios where the OBD varies across responsive indications, such as scenario 8. In scenario 8, $d_H$ is the true OBD for $I_2,$ while $d_L$ is the true OBD for $I_3$ and $I_4$. The CSPs of ROMI-v1 and ROMI-v2 are 4\% and 2.6\% lower than the Independent design but outperform ROMI-v1-NC, with CSP improvements of 6.4\% and 7.8\%, respectively. This shows the benefit of clustering indications under the Bayesian hierarchical model in ROMI-v1 and ROMI-v2. The Pool design has the lowest CSP, about 56.6\%, and the highest probability of selecting an ineffective dose for $I_1$.

The advantage of information borrowing increases with the number of responsive indications, shown by scenarios 9, 10, and 11, where all indications respond to treatment, resulting in comparable sample sizes across all designs. In homogeneous scenarios where OBDs are consistent across indications, the Pool and ROMI designs have higher CSPs than the Independent design. For example, in scenario 9, ROMI-v1 shows a 2.4\% increase in CSP, ROMI-v2 a 5.7\% increase, and ROMI-v1-NC a 9\% increase, compared to Independent. The Pool design has the highest CSP of 81.8\%, essentially because its homogeneity assumption happens to be correct in this scenario.
In the heterogeneous$^2$ scenarios, the OBD varies across responsive indications. For example, in scenario 11, $d_H$ is the true OBD for $I_1$ and $I_2$, and $d_L$ is the true OBD for $I_3$ and $I_4$. ROMI-v1 and ROMI-v2 have CSPs about 5\% lower than the Independent design, but outperform ROMI-v1-NC by 9\%. The Pool design shows the poorest performance, correctly selecting the OBD with only a 50\% CSP.

\subsection{Sensitivity analysis}\label{subsec:sensitivity}
We examined the performance of ROMI-v1 and ROMI-v2 in the presence of $\pi_{R,H,k}$ drift for $d_H$ between stages, exploring the impacts of both positive and negative drifts. Table \ref{tab:sim_results_eff_drift} gives simulation results where $\pi_{R,H,k}$ increased by 0.025 from stage 1 to stage 2 in the upper portion of the table, and decreased by 0.025 in the lower portion. This increment corresponds to 25\% of the maximum $\pi_{R,H,k} - \pi_{R,L,k}$ difference of .10 in our simulation settings. In each of scenarios 9 - 11, all $I_k$'s are responsive to both $d_H$ and $d_L$. Compared to ROMI-v1,  ROMI-v2 demonstrates similar or better accuracy in selecting OBD across all scenarios. Thus, ROMI-v2 does a good job of handling drift in response rates between stages.

\begin{figure}[H]
    \centering
    \includegraphics[width=1\textwidth]{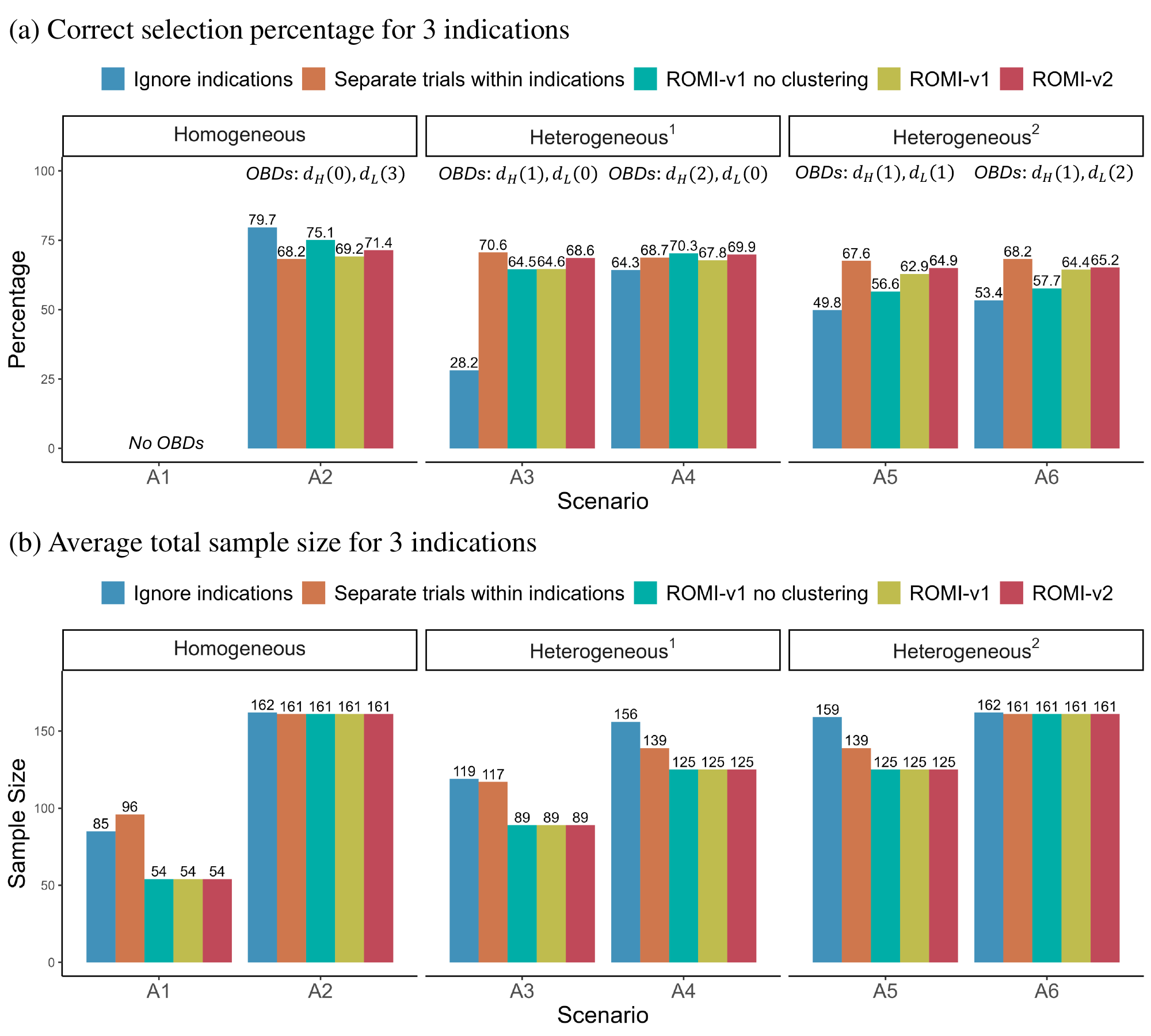}
    \caption{{\bf For three indications}, (a) Correct selection percentages and (b) average total sample sizes for the Pool design that ignores indications, Independent design that conducts separate trials within indications, ROMI-v1 with no indication clustering, ROMI-v1 with clustering, and ROMI-v2 with clustering.
    In homogeneous scenarios,  OBDs are identical across indications. Heterogeneous$^1$ scenarios include some non-responsive indications and identical OBDs among responsive indications. In Heterogeneous$^2$ scenarios,   OBDs vary among responsive indications. }
    \label{fig:sim-3-ind}
\end{figure}

\begin{figure}[H]
    \centering
    \includegraphics[width=1\textwidth]{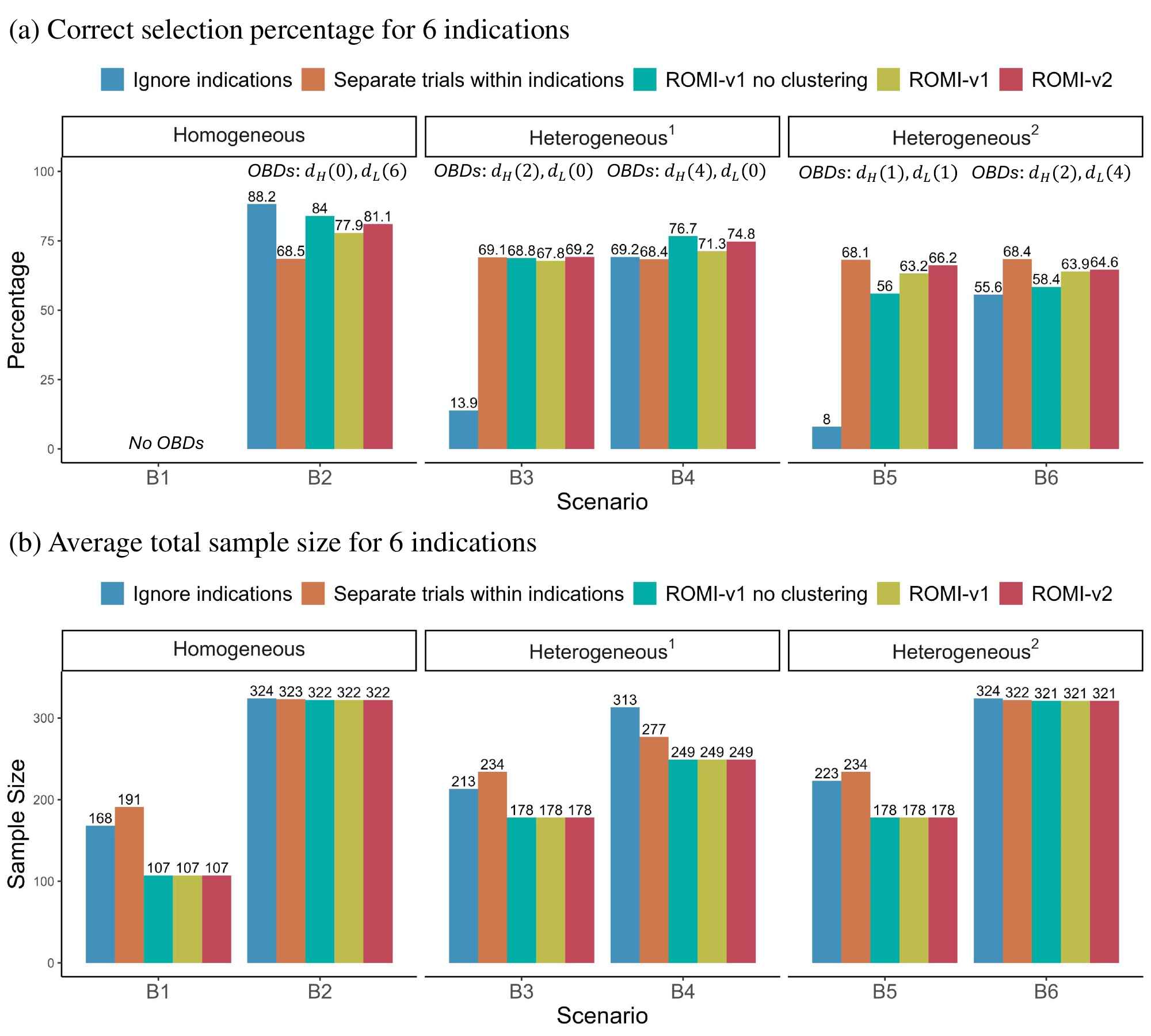}
    \caption{{\bf For six indications}, (a) correct selection percentages and (b) average total sample sizes for the Pool design that ignores indications, Independent design that conducts separate trials within indications, ROMI-v1 with no indication clustering, ROMI-v1 with clustering, and ROMI-v2 with clustering.
    In homogeneous scenarios,  OBDs are identical across indications. Heterogeneous$^1$ scenarios include some non-responsive indications and identical OBDs among responsive indications. In Heterogeneous$^2$ scenarios,   OBDs vary among responsive indications.}
    \label{fig:sim-6-ind}
\end{figure}

We also evaluated the performance of the ROMI designs for a trial with either $K=3$ or $K=6$ indications, illustrated in  Figures \ref{fig:sim-3-ind} and \ref{fig:sim-6-ind}.  The ROMI designs reduce sample size compared to the Pool and Independent designs when some $I_k$'s are non-responsive to treatment.  As expected, the Pool design has the highest CSP when all indications have the same dose-outcome curves but performs very poorly when the dose-outcome curves vary across indications. ROMI-v2 shows similar or superior OBD selection compared to ROMI-v1. For trials with $K=3$ indications, ROMI-v2 is comparable to the Independent design and outperforms ROMI-v1-NC when OBDs vary across responsive indications in accurately selecting OBDs. The performance of ROMI-v1 and ROMI-v2 improves as the number of indications increases. In Scenario B2, where $K=6$ indications are responsive, the CSP values of the ROMI-v1 and ROMI-v2 designs are 9.4\% and 12.6\% higher than the Independent design, respectively. Detailed results are provided in Supplementary Material S2.

As a final sensitivity analysis, we evaluated the ROMI designs, assuming that the shrinkage parameter follows a Half-Cauchy distribution. Simulation results are given in Supplementary Material S3. While this provides greater robustness, it reduces information borrowing.

\section{Discussion}  \label{sec:Discussion}

ROMI effectively identifies and discontinues indications not responsive to treatment,  substantially reducing sample size compared to designs that ignore indications or optimize dose independently for each indication. 
When dose-outcome curves differ between indications, ROMI  accurately identifies indication-specific OBDs. The version of ROMI that uses information from both stages shows similar or higher accuracy in OBD selection compared to the version that ignores stage 1 data on $d_H$. Compared to conducting separate trials within indications, the second version of ROMI has greater accuracy in identifying the OBD if it is the same across indications. When the OBDs vary across indications, the accuracy of the ROMI design is slightly lower than the Independent design, but it still outperforms the design with ROMI structure but does not cluster similar indications. For a larger number of indications, the performance of the ROMI design improves.


As a future study, it may be worthwhile to develop a Bayesian hierarchical model accounting for count variables $\boldsymbol{X}_{\ell,k}$. Stage 1 screening of ROMI is based on the assumption that $d_H$ cannot be less effective than $d_L$. If this is invalid, stage 1 can be removed, with randomization for all indications throughout. In addition to efficacy and toxicity, endpoints such as pharmacokinetics or quality of life,  may be included in the final OBD selection.

\clearpage
\bibliographystyle{plainnat}


\newpage
\begin{table}[h!]
\centering
\caption{Example of indication-specific utilities for two binary outcomes}
\begin{tabular}{ccc}
 \multicolumn{3}{c}{\bf Indication 1}  \\
                  & $Y_R=1$           & $Y_R=0$           \\
                  \hline
         $Y_T=0$  & $U_1(0,1)=100$          & $U_1(0,0)=40$         \\
        $Y_T=1$ & $U_1(1,1)=60$          & $U_1(1,0)=0$          \\
        \hline
        \\
  \multicolumn{3}{c}{\bf Indication 2}  \\
                  & $Y_R=1$           & $Y_R=0$           \\
                  \hline
         $Y_T=0$  & $U_2(0,1)=100$          & $U_2(0,0)=20$         \\
        $Y_T=1$ & $U_2(1,1)=80$          & $U_2(1,0)=0$          \\ \hline
                \\
  \multicolumn{3}{c}{\bf Indication 3}  \\
                  & $Y_R=1$           & $Y_R=0$           \\
                  \hline
         $Y_T=0$  & $U_3(0,1)=100$          & $U_3(0,0)=60$         \\
        $Y_T=1$ & $U_3(1,1)= 40$          & $U_3(1,0)=0$          \\
        \hline
        \\
   \multicolumn{3}{c}{\bf Indication 4}  \\
                  & $Y_R=1$           & $Y_R=0$           \\
                  \hline
         $Y_T=0$  & $U_4(0,1)=100$          & $U_4(0,0)=30$         \\
        $Y_T=1$ & $U_4(1,1)= 70$          & $U_4(1,0)=0$          \\
        \hline
\end{tabular}
\label{tab:utility-example}
\end{table}

\newpage
\begin{longtable}{llllllllllll}
\caption{Simulation results for the Pool, Independent, ROMI-v1-NC, ROMI-v1, and ROMI-v2 designs. Values for the true OBD of each indication are given in boldface. Doses are indexed by  $\ell = L, H$ and indications by $\ k=1,2,3,4.$ }
\label{tab:simulation-results}\\
\hline
 &  & \multicolumn{8}{c}{Percent  selection of the dose as OBD} &  &  \\ \cline{3-10}
 &  & \multicolumn{2}{c}{$I_1$} & \multicolumn{2}{c}{$I_2$} & \multicolumn{2}{c}{$I_3$} & \multicolumn{2}{c}{$I_4$} &  &  \\ \cline{3-10}
\multirow{-3}{*}{Design} & \multirow{-3}{*}{\textbf{}} & $d_H$ & $d_L$ & $d_H$ & $d_L$ & $d_H$ & $d_L$ & $d_H$ & $d_L$ & \multirow{-3}{*}{CSP} & \multirow{-3}{*}{$N$} \\ \hline
\endfirsthead
\multicolumn{12}{c}%
{{\bfseries Table \thetable\ continued from previous page}} \\
\hline
 &  & \multicolumn{8}{c}{Percent  selection of the dose as OBD} &  &  \\ \cline{3-10}
 &  & \multicolumn{2}{c}{$I_1$} & \multicolumn{2}{c}{$I_2$} & \multicolumn{2}{c}{$I_3$} & \multicolumn{2}{c}{$I_4$} &  &  \\ \cline{3-10}
\multirow{-3}{*}{Design} & \multirow{-3}{*}{\textbf{}} & $d_H$ & $d_L$ & $d_H$ & $d_L$ & $d_H$ & $d_L$ & $d_H$ & $d_L$ & \multirow{-3}{*}{CSP} & \multirow{-3}{*}{$N$} \\ \hline
\endhead
\rowcolor[HTML]{EFEFEF}
Scenario 1 &  &  &  &  &  &  &  &  &  &  &  \\
\rowcolor[HTML]{EFEFEF}
 & $\pi_{T,\ell,k}^{true}$ & 0.40 & 0.30 & 0.40 & 0.30 & 0.40 & 0.30 & 0.40 & 0.30 &  &  \\
\rowcolor[HTML]{EFEFEF}
 & $\pi_{R,\ell,k}^{true}$ & 0.05 & 0.05 & 0.05 & 0.05 & 0.05 & 0.05 & 0.05 & 0.05 &  &  \\
\rowcolor[HTML]{EFEFEF}
 & $\bar{U}_{\ell,k}^{true}$ & 27 & 31 & 27 & 31 & 27 & 31 & 27 & 31 &  &  \\
\rowcolor[HTML]{EFEFEF}
Pool &  & 0.0 & 0.0 & 0.0 & 0.0 & 0.0 & 0.0 & 0.0 & 0.0 &NA  & 113 \\
\rowcolor[HTML]{EFEFEF}
Independent &  & 2.1 & 3.3 & 2.5 & 4.0 & 2.5 & 2.8 & 2.9 & 3.3 &NA  & 127 \\
\rowcolor[HTML]{EFEFEF}
ROMI-v1-NC &  & 1.3 & 0.9 & 0.5 & 0.6 & 0.6 & 0.7 & 1.3 & 0.7 &NA  & 71 \\
\rowcolor[HTML]{EFEFEF}
ROMI-v1 &  & 1.3 & 0.9 & 0.5 & 0.6 & 0.6 & 0.7 & 1.3 & 0.7 &NA  & 71 \\
\rowcolor[HTML]{EFEFEF}
ROMI-v2 &  & 1.3 & 0.9 & 0.5 & 0.7 & 0.7 & 0.8 & 1.3 & 0.8 & NA & 71 \\
Scenario 2 &  &  &  &  &  &  &  &  &  &  &  \\
 & $\pi_{T,\ell,k}^{true}$ & \textbf{0.2} & 0.15 & 0.40 & 0.30 & 0.40 & 0.30 & 0.40 & 0.30 &  &  \\
 & $\pi_{R,\ell,k}^{true}$ & \textbf{0.4} & 0.3 & 0.05 & 0.05 & 0.05 & 0.05 & 0.05 & 0.05 &  &  \\
 & $\bar{U}_{\ell,k}^{true}$ & \textbf{56} & 52 & 27 & 31 & 27 & 31 & 27 & 31 &  &  \\
Pool &  & \textbf{5.5} & 0.8 & 5.5 & 0.8 & 5.5 & 0.8 & 5.5 & 0.8 & 5.5 & 128 \\
Independent &  & \textbf{69.8} & 30.2 & 2.6 & 3.1 & 2.8 & 2.8 & 3.0 & 3.3 & 69.8 & 149 \\
ROMI-v1-NC &  & \textbf{64.5} & 35.0 & 0.7 & 1.0 & 0.9 & 0.8 & 1.0 & 0.7 & 64.5 & 107 \\
ROMI-v1 &  & \textbf{65.0} & 34.4 & 0.7 & 1.0 & 0.9 & 0.8 & 1.0 & 0.7 & 65.0 & 107 \\
ROMI-v2 &  & \textbf{69.7} & 29.8 & 0.7 & 1.1 & 0.9 & 0.8 & 1.0 & 0.8 & 69.7 & 107 \\
\rowcolor[HTML]{EFEFEF}
Scenario 3 &  &  &  &  &  &  &  &  &  &  &  \\
\rowcolor[HTML]{EFEFEF}
 & $\pi_{T,\ell,k}^{true}$ & 0.25 & \textbf{0.15} & 0.40 & 0.30 & 0.40 & 0.30 & 0.40 & 0.30 &  &  \\
\rowcolor[HTML]{EFEFEF}
 & $\pi_{R,\ell,k}^{true}$ & 0.4 & \textbf{0.40} & 0.05 & 0.05 & 0.05 & 0.05 & 0.05 & 0.05 &  &  \\
\rowcolor[HTML]{EFEFEF}
 & $\bar{U}_{\ell,k}^{true}$ & 54 & \textbf{58} & 27 & 31 & 27 & 31 & 27 & 31 &  &  \\
\rowcolor[HTML]{EFEFEF}
\cellcolor[HTML]{EFEFEF}Pool &  & 4.0 & \textbf{4.3} & 4.0 & 4.3 & 4.0 & 4.3 & 4.0 & 4.3 & 4.3 & 135 \\
\rowcolor[HTML]{EFEFEF}
\cellcolor[HTML]{EFEFEF}Independent &  & 32.1 & \textbf{68.0} & 2.4 & 3.3 & 2.7 & 3.3 & 2.6 & 3.1 & 68.0 & 149 \\
\rowcolor[HTML]{EFEFEF}
\cellcolor[HTML]{EFEFEF}ROMI-v1-NC &  & 30.6 & \textbf{68.2} & 0.9 & 0.5 & 0.9 & 0.6 & 1.1 & 1.0 & 68.2 & 107 \\
\rowcolor[HTML]{EFEFEF}
\cellcolor[HTML]{EFEFEF}ROMI-v1 &  & 30.6 & \textbf{68.3} & 0.9 & 0.5 & 0.9 & 0.6 & 1.1 & 1.0 & 68.3 & 107 \\
\rowcolor[HTML]{EFEFEF}
\cellcolor[HTML]{EFEFEF}ROMI-v2 &  & 30.0 & \textbf{68.9} & 0.9 & 0.6 & 0.9 & 0.7 & 1.2 & 1.0 & 68.9 & 107 \\
Scenario 4 &  &  &  &  &  &  &  &  &  &  &  \\
 & $\pi_{T,\ell,k}^{true}$ & \textbf{0.2} & 0.15 & 0.40 & 0.30 & 0.40 & 0.30 & \textbf{0.20} & 0.15 &  &  \\
 & $\pi_{R,\ell,k}^{true}$ & \textbf{0.4} & 0.3 & 0.05 & 0.05 & 0.05 & 0.05 & \textbf{0.40} & 0.30 &  &  \\
 & $\bar{U}_{\ell,k}^{true}$ & \textbf{56} & 52 & 27 & 31 & 27 & 31 & \textbf{56} & 52 &  &  \\
Pool &  & \textbf{63.0} & 24.4 & 63.0 & 24.4 & 63.0 & 24.4 & \textbf{63.0} & 24.4 & 63.0 & 185 \\
Independent &  & \textbf{68.7} & 31.3 & 2.4 & 3.2 & 2.7 & 3.3 & \textbf{69.3} & 30.7 & 69.0 & 170 \\
ROMI-v1-NC &  & \textbf{72.6} & 27.0 & 0.7 & 0.8 & 1.0 & 1.1 & \textbf{70.6} & 28.7 & 71.6 & 143 \\
ROMI-v1 &  & \textbf{70.3} & 29.2 & 0.7 & 0.8 & 1.0 & 1.1 & \textbf{66.5} & 32.7 & 68.4 & 143 \\
ROMI-v2 &  & \textbf{72.6} & 26.9 & 0.8 & 0.8 & 1.0 & 1.1 & \textbf{68.8} & 30.5 & 70.7 & 143 \\
\rowcolor[HTML]{EFEFEF}
Scenario 5 &  &  &  &  &  &  &  &  &  &  &  \\
\rowcolor[HTML]{EFEFEF}
 & $\pi_{T,\ell,k}^{true}$ & 0.25 & \textbf{0.15} & 0.40 & 0.30 & 0.40 & 0.30 & 0.25 & \textbf{0.15} &  &  \\
\rowcolor[HTML]{EFEFEF}
 & $\pi_{R,\ell,k}^{true}$ & 0.4 & \textbf{0.4} & 0.05 & 0.05 & 0.05 & 0.05 & 0.40 & \textbf{0.40} &  &  \\
\rowcolor[HTML]{EFEFEF}
 & $\bar{U}_{\ell,k}^{true}$ & 54 & \textbf{58} & 27 & 31 & 27 & 31 & 54 & \textbf{58} &  &  \\
\rowcolor[HTML]{EFEFEF}
\cellcolor[HTML]{EFEFEF}Pool &  & 24.4 & \textbf{71.7} & 24.4 & 71.7 & 24.4 & 71.7 & 24.4 & \textbf{71.7} & 71.7 & 202 \\
\rowcolor[HTML]{EFEFEF}
\cellcolor[HTML]{EFEFEF}Independent &  & 32.1 & \textbf{67.9} & 2.4 & 2.9 & 2.7 & 3.3 & 30.3 & \textbf{69.8} & 68.8 & 171 \\
\rowcolor[HTML]{EFEFEF}
\cellcolor[HTML]{EFEFEF}ROMI-v1-NC &  & 27.2 & \textbf{71.7} & 0.5 & 1.2 & 1.1 & 0.9 & 27.4 & \textbf{71.6} & 71.7 & 143 \\
\rowcolor[HTML]{EFEFEF}
\cellcolor[HTML]{EFEFEF}ROMI-v1 &  & 31.4 & \textbf{67.4} & 0.5 & 1.2 & 1.1 & 0.9 & 29.9 & \textbf{69.2} & 68.3 & 143 \\
\rowcolor[HTML]{EFEFEF}
\cellcolor[HTML]{EFEFEF}ROMI-v2 &  & 31.0 & \textbf{68.0} & 0.5 & 1.2 & 1.1 & 0.9 & 28.6 & \textbf{70.5} & 69.2 & 143 \\
Scenario 6 &  &  &  &  &  &  &  &  &  &  &  \\
 & $\pi_{T,\ell,k}^{true}$ & 0.40 & 0.30 & \textbf{0.20} & 0.15 & \textbf{0.20} & 0.15 & \textbf{0.20} & 0.15 &  &  \\
 & $\pi_{R,\ell,k}^{true}$ & 0.05 & 0.05 & \textbf{0.40} & 0.30 & \textbf{0.40} & 0.30 & \textbf{0.40} & 0.30 &  &  \\
 & $\bar{U}_{\ell,k}^{true}$ & 27 & 31 & \textbf{56} & 52 & \textbf{56} & 52 & \textbf{56} & 52 &  &  \\
Pool &  & 71.4 & 28.6 & \textbf{71.4} & 28.6 & \textbf{71.4} & 28.6 & \textbf{71.4} & 28.6 & 71.4 & 210 \\
Independent &  & 3.3 & 3.2 & \textbf{70.1} & 29.9 & \textbf{67.9} & 32.1 & \textbf{68.2} & 31.9 & 68.7 & 192 \\
ROMI-v1-NC &  & 1.2 & 1.0 & \textbf{74.0} & 25.4 & \textbf{73.9} & 25.2 & \textbf{73.2} & 26.0 & 73.7 & 178 \\
ROMI-v1 &  & 1.2 & 1.0 & \textbf{69.6} & 29.8 & \textbf{68.0} & 31.0 & \textbf{67.8} & 31.3 & 68.5 & 178 \\
ROMI-v2 &  & 1.2 & 1.0 & \textbf{72.0} & 27.4 & \textbf{70.8} & 28.3 & \textbf{71.1} & 28.1 & 71.3 & 178 \\
\rowcolor[HTML]{EFEFEF}
Scenario 7 &  &  &  & \textbf{} &  & \textbf{} &  & \textbf{} &  &  &  \\
\rowcolor[HTML]{EFEFEF}
 & $\pi_{T,\ell,k}^{true}$ & 0.40 & 0.30 & 0.25 & \textbf{0.15} & 0.25 & \textbf{0.15} & 0.25 & \textbf{0.15} &  &  \\
\rowcolor[HTML]{EFEFEF}
 & $\pi_{R,\ell,k}^{true}$ & 0.05 & 0.05 & 0.40 & \textbf{0.40} & 0.40 & \textbf{0.40} & 0.40 & \textbf{0.40} &  &  \\
\rowcolor[HTML]{EFEFEF}
 & $\bar{U}_{\ell,k}^{true}$ & 27 & 31 & 54 & \textbf{58} & 54 & \textbf{58} & 54 & \textbf{58} &  &  \\
\rowcolor[HTML]{EFEFEF}
\cellcolor[HTML]{EFEFEF}Pool &  & 14.8 & 85.3 & 14.8 & \textbf{85.3} & 14.8 & \textbf{85.3} & 14.8 & \textbf{85.3} & 85.3 & 216 \\
\rowcolor[HTML]{EFEFEF}
\cellcolor[HTML]{EFEFEF}Independent &  & 2.7 & 3.4 & 31.8 & \textbf{68.2} & 32.0 & \textbf{68.0} & 31.2 & \textbf{68.9} & 68.4 & 194 \\
\rowcolor[HTML]{EFEFEF}
\cellcolor[HTML]{EFEFEF}ROMI-v1-NC &  & 1.3 & 1.0 & 25.4 & \textbf{73.9} & 24.9 & \textbf{74.0} & 23.9 & \textbf{75.2} & 74.4 & 179 \\
\rowcolor[HTML]{EFEFEF}
\cellcolor[HTML]{EFEFEF}ROMI-v1 &  & 1.3 & 1.0 & 31.2 & \textbf{68.0} & 30.2 & \textbf{68.6} & 28.0 & \textbf{71.2} & 69.3 & 179 \\
\rowcolor[HTML]{EFEFEF}
\cellcolor[HTML]{EFEFEF}ROMI-v2 &  & 1.3 & 0.9 & 28.6 & \textbf{70.7} & 29.0 & \textbf{69.9} & 26.0 & \textbf{73.2} & 71.2 & 179 \\
Scenario 8 &  &  &  &  & \textbf{} &  & \textbf{} &  & \textbf{} &  &  \\
 & $\pi_{T,\ell,k}^{true}$ & 0.40 & 0.30 & \textbf{0.20} & 0.15 & 0.25 & \textbf{0.15} & 0.25 & \textbf{0.15} &  &  \\
 & $\pi_{R,\ell,k}^{true}$ & 0.05 & 0.05 & \textbf{0.40} & 0.30 & 0.4 & \textbf{0.40} & 0.40 & \textbf{0.40} &  & \textbf{} \\
 & $\bar{U}_{\ell,k}^{true}$ & 27 & 31 & \textbf{56} & 52 & 54 & \textbf{58} & 54 & \textbf{58} &  &  \\
Pool &  & 30.4 & 69.7 & \textbf{30.4} & 69.7 & 30.4 & \textbf{69.7} & 30.4 & \textbf{69.7} & 56.6 & 215 \\
Independent &  & 2.7 & 3.6 & \textbf{68.5} & 31.5 & 31.3 & \textbf{68.7} & 32.2 & \textbf{67.8} & 68.3 & 193 \\
ROMI-v1-NC &  & 1.4 & 1.0 & \textbf{47.7} & 51.6 & 36.2 & \textbf{62.7} & 35.8 & \textbf{63.3} & 57.9 & 179 \\
ROMI-v1 &  & 1.4 & 1.0 & \textbf{61.4} & 38.0 & 33.7 & \textbf{65.2} & 33.0 & \textbf{66.2} & 64.3 & 179 \\
ROMI-v2 &  & 1.4 & 1.0 & \textbf{62.5} & 36.9 & 32.3 & \textbf{66.7} & 31.1 & \textbf{68.1} & 65.7 & 179 \\
\rowcolor[HTML]{EFEFEF}
Scenario 9 &  &  &  & \textbf{} &  &  & \textbf{} &  & \textbf{} &  &  \\
\rowcolor[HTML]{EFEFEF}
 & $\pi_{T,\ell,k}^{true}$ & \textbf{0.2} & 0.15 & \textbf{0.2} & 0.15 & \textbf{0.2} & 0.15 & \textbf{0.2} & 0.15 &  &  \\
\rowcolor[HTML]{EFEFEF}
 & $\pi_{R,\ell,k}^{true}$ & \textbf{0.4} & 0.3 & \textbf{0.4} & 0.3 & \textbf{0.4} & 0.3 & \textbf{0.4} & 0.3 &  &  \\
\rowcolor[HTML]{EFEFEF}
 & $\bar{U}_{\ell,k}^{true}$ & \textbf{56} & 52 & \textbf{56} & 52 & \textbf{56} & 52 & \textbf{56} & 52 &  &  \\
\rowcolor[HTML]{EFEFEF}
\cellcolor[HTML]{EFEFEF}Pool &  & \textbf{81.8} & 18.2 & \textbf{81.8} & 18.2 & \textbf{81.8} & 18.2 & \textbf{81.8} & 18.2 & 81.8 & 216 \\
\rowcolor[HTML]{EFEFEF}
\cellcolor[HTML]{EFEFEF}Independent &  & \textbf{69.9} & 30.1 & \textbf{69.4} & 30.6 & \textbf{68.6} & 31.5 & \textbf{67.3} & 32.7 & 68.8 & 214 \\
\rowcolor[HTML]{EFEFEF}
\cellcolor[HTML]{EFEFEF}ROMI-v1-NC &  & \textbf{77.8} & 21.6 & \textbf{78.3} & 21.1 & \textbf{77.7} & 21.4 & \textbf{77.4} & 21.8 & 77.8 & 214 \\
\rowcolor[HTML]{EFEFEF}
\cellcolor[HTML]{EFEFEF}ROMI-v1 &  & \textbf{71.7} & 27.8 & \textbf{70.9} & 28.6 & \textbf{70.8} & 28.4 & \textbf{71.2} & 28.0 & 71.2 & 214 \\
\rowcolor[HTML]{EFEFEF}
\cellcolor[HTML]{EFEFEF}ROMI-v2 &  & \textbf{75.1} & 24.4 & \textbf{74.4} & 25.1 & \textbf{74.4} & 24.7 & \textbf{74.0} & 25.2 & 74.5 & 214 \\
Scenario 10 &  & \textbf{} &  & \textbf{} &  & \textbf{} &  & \textbf{} &  &  &  \\
 & $\pi_{T,\ell,k}^{true}$ & 0.25 & \textbf{0.15} & 0.25 & \textbf{0.15} & 0.25 & \textbf{0.15} & 0.25 & \textbf{0.15} &  &  \\
 & $\pi_{R,\ell,k}^{true}$ & 0.40 & \textbf{0.40} & 0.40 & \textbf{0.40} & 0.40 & \textbf{0.40} & 0.4 & \textbf{0.40} &  &  \\
 & $\bar{U}_{\ell,k}^{true}$ & 54 & \textbf{58} & 54 & \textbf{58} & 54 & \textbf{58} & 54 & \textbf{58} &  &  \\
Pool &  & 16.9 & \textbf{83.1} & 16.9 & \textbf{83.1} & 16.9 & \textbf{83.1} & 16.9 & \textbf{83.1} & 83.1 & 216 \\
Independent &  & 33.0 & \textbf{67.0} & 31.6 & \textbf{68.4} & 33.2 & \textbf{66.8} & 31.3 & \textbf{68.7} & 67.7 & 216 \\
ROMI-v1-NC &  & 21.3 & \textbf{77.6} & 21.3 & \textbf{77.8} & 22.0 & \textbf{76.8} & 22.2 & \textbf{76.9} & 77.3 & 214 \\
ROMI-v1 &  & 27.0 & \textbf{71.9} & 27.0 & \textbf{72.2} & 29.0 & \textbf{69.8} & 27.0 & \textbf{72.2} & 71.5 & 214 \\
ROMI-v2 &  & 26.0 & \textbf{72.9} & 24.0 & \textbf{75.2} & 26.5 & \textbf{72.4} & 24.3 & \textbf{74.9} & 73.8 & 214 \\
\rowcolor[HTML]{EFEFEF}
Scenario 11 &  & \textbf{} &  &  & \textbf{} &  & \textbf{} &  & \textbf{} &  &  \\
\rowcolor[HTML]{EFEFEF}
 & $\pi_{T,\ell,k}^{true}$ & \textbf{0.20} & 0.15 & \textbf{0.20} & 0.15 & 0.25 & \textbf{0.15} & 0.25 & \textbf{0.15} &  &  \\
 \rowcolor[HTML]{EFEFEF}
 & $\pi_{R,\ell,k}^{true}$ & \textbf{0.40} & 0.30 & \textbf{0.40} & 0.30 & 0.40 & \textbf{0.40} & 0.40 & \textbf{0.40} &  &  \\
 \rowcolor[HTML]{EFEFEF}
 & $\bar{U}_{\ell,k}^{true}$ & \textbf{56} & 52 & \textbf{56} & 52 & 54 & \textbf{58} & 54 & \textbf{58} &  &  \\
 \rowcolor[HTML]{EFEFEF}
 Pool &  & \textbf{50.9} & 49.1 & \textbf{50.9} & 49.1 & 50.9 & \textbf{49.1} & 50.9 & \textbf{49.1} & 50.0 & 216 \\
 \rowcolor[HTML]{EFEFEF}
Independent &  & \textbf{69.2} & 30.8 & \textbf{68.9} & 31.1 & 30.2 & \textbf{69.8} & 31.6 & \textbf{68.5} & 69.1 & 215 \\
\rowcolor[HTML]{EFEFEF}
 ROMI-v1-NC&  & \textbf{54.9} & 44.4 & \textbf{54.3} & 45.1 & 43.8 & \textbf{55.1} & 42.4 & \textbf{56.7} & 55.3 & 214 \\
\rowcolor[HTML]{EFEFEF}
 ROMI-v1&  & \textbf{63.4} & 36.0 & \textbf{62.1} & 37.4 & 36.0 & \textbf{62.9} & 36.0 & \textbf{63.1} & 62.9 & 214 \\
\rowcolor[HTML]{EFEFEF}
 ROMI-v2 & & \textbf{64.3} & 35.1 & \textbf{64.2} & 35.2 & 35.6 & \textbf{63.4} & 35.2 & \textbf{64.0} & 64.0 & 214 \\ \hline
\end{longtable}
\vspace{-2em}
 \begin{tablenotes}
 \begin{centering}
      \small
      \item Abbreviations: CSP: correct selection percentage; $N$: average total sample size.
\end{centering}
    \end{tablenotes}

\newpage

\begin{longtable}{lllllllllll}
\caption{Sensitivity analysis of ROMI designs with efficacy drift of $d_H$ effects between stage 1 and stage 2. Values for the true OBD of each indication are given in boldface. Doses are indexed by  $\ell = L, H$ and indications by $\ k=1,2,3,4.$ }
\label{tab:sim_results_eff_drift}\\
\hline
&  & \multicolumn{8}{c}{Percent  selection of the dose as OBD} &   \\ \cline{3-10}
\multicolumn{2}{c}{} & \multicolumn{2}{c}{$I_1$} & \multicolumn{2}{c}{$I_2$} & \multicolumn{2}{c}{$I_3$} & \multicolumn{2}{c}{$I_4$} & \multicolumn{1}{c}{} \\ \cline{3-10}
\multicolumn{2}{c}{\multirow{-3}{*}{Design}} & $d_H$ & $d_L$ & $d_H$ & $d_L$ & $d_H$ & $d_L$ & $d_H$ & $d_L$ & \multicolumn{1}{c}{\multirow{-3}{*}{CSP}} \\ \hline
\endfirsthead
\multicolumn{11}{c}%
{{\bfseries Table \thetable\ continued from previous page}} \\
\hline
\multicolumn{2}{c}{} & \multicolumn{2}{l}{$I_1$} & \multicolumn{2}{l}{$I_2$} & \multicolumn{2}{l}{$I_3$} & \multicolumn{2}{l}{$I_4$} & \multicolumn{1}{c}{} \\ \cline{3-10}
\multicolumn{2}{c}{\multirow{-2}{*}{Design}} & $d_H$ & $d_L$ & $d_H$ & $d_L$ & $d_H$ & $d_L$ & $d_H$ & $d_L$ & \multicolumn{1}{c}{\multirow{-2}{*}{CSP}} \\ \hline
\endhead
\hline
\endfoot
\endlastfoot
\\
\multicolumn{11}{c}{\textit{Positive Drift}} \\
\rowcolor[HTML]{EFEFEF}
\multicolumn{2}{l}{\cellcolor[HTML]{EFEFEF}Scenario 9} &  &  &  &  &  &  &  &  &  \\
\rowcolor[HTML]{EFEFEF}
\multicolumn{2}{l}{\cellcolor[HTML]{EFEFEF}ROMI-v1} & \textbf{72.0} & 27.0 & \textbf{71.8} & 27.0 & \textbf{71.3} & 27.2 & \textbf{72.1} & 26.6 & 71.8 \\
\rowcolor[HTML]{EFEFEF}
\multicolumn{2}{l}{\cellcolor[HTML]{EFEFEF}ROMI-v2} & \textbf{72.1} & 26.9 & \textbf{71.5} & 27.4 & \textbf{72.1} & 26.4 & \textbf{71.7} & 27.0 & 71.8 \\
\multicolumn{2}{l}{Scenario 10} &  &  &  &  &  &  &  &  &  \\
\multicolumn{2}{l}{ROMI-v1} & 28.3 & \textbf{70.3} & 27.2 & \textbf{71.7} & 27.4 & \textbf{70.7} & 27.0 & \textbf{71.3} & 71.0 \\
\multicolumn{2}{l}{ROMI-v2} & 23.6 & \textbf{75.0} & 22.2 & \textbf{76.8} & 22.8 & \textbf{75.2} & 21.9 & \textbf{76.4} & 75.8 \\
\rowcolor[HTML]{EFEFEF}
\rowcolor[HTML]{EFEFEF}
\multicolumn{2}{l}{Scenario 11} &  &  &  &  &  &  &  &  &  \\
\rowcolor[HTML]{EFEFEF}
\multicolumn{2}{l}{ROMI-v1} & \textbf{62.8} & 36.0 & \textbf{61.6} & 37.2 & 35.4 & \textbf{62.7} & 36.6 & \textbf{61.7} & 62.2 \\
\rowcolor[HTML]{EFEFEF}
\multicolumn{2}{l}{ROMI-v2} & \textbf{60.6} & 38.3 & \textbf{59.9} & 38.9 & 31.7 & \textbf{66.5} & 31.5 & \textbf{66.8} & 63.4 \\ \hline
\\
\multicolumn{11}{c}{\textit{Negative Drift}} \\
\rowcolor[HTML]{EFEFEF}
\multicolumn{2}{l}{\cellcolor[HTML]{EFEFEF}Scenario 9} &  &  &  &  &  &  &  &  &  \\
\rowcolor[HTML]{EFEFEF}
\multicolumn{2}{l}{\cellcolor[HTML]{EFEFEF}ROMI-v1} & \textbf{71.8} & 27.8 & \textbf{71.4} & 28.1 & \textbf{71.7} & 27.8 & \textbf{71.9} & 27.5 & 71.7 \\
\rowcolor[HTML]{EFEFEF}
\multicolumn{2}{l}{\cellcolor[HTML]{EFEFEF}ROMI-v2} & \textbf{77.3} & 22.4 & \textbf{77.4} & 22.1 & \textbf{78.3} & 21.2 & \textbf{77.4} & 22.0 & 77.6 \\
\multicolumn{2}{l}{Scenario 10} &  &  &  &  &  &  &  &  &  \\
\multicolumn{2}{l}{ROMI-v1} & 27.4 & \textbf{71.9} & 26.4 & \textbf{72.8} & 28.4 & \textbf{71.0} & 25.6 & \textbf{73.9} & 72.4 \\
\multicolumn{2}{l}{ROMI-v2} & 28.2 & \textbf{71.1} & 25.1 & \textbf{74.2} & 27.4 & \textbf{72.0} & 25.4 &\textbf{ 74.1} & 72.8 \\
\rowcolor[HTML]{EFEFEF}
\multicolumn{2}{l}{Scenario 11} &  &  &  &  &  &  &  &  &  \\
\rowcolor[HTML]{EFEFEF}
\multicolumn{2}{l}{ROMI-v1} & \textbf{64.9} & 34.6 & \textbf{62.2} & 37.4 & 36.1 & \textbf{63.2} & 36.8 & \textbf{62.6} & 63.2 \\
\rowcolor[HTML]{EFEFEF}
\multicolumn{2}{l}{ROMI-v2} & \textbf{68.5} & 31.1 & \textbf{67.0} & 32.6 & 39.5 & \textbf{59.9} & 38.9 & \textbf{60.6} & 64.0 \\ \hline
\end{longtable}
\vspace{-2em}
 \begin{tablenotes}
 \begin{centering}
      \small
      \item \hspace{2em} Abbreviations: CSP: correct selection percentage.
\end{centering}
    \end{tablenotes}

\clearpage\pagebreak

\baselineskip=24pt
\begin{center}
{\Large \bf Supplementary Materials for ROMI:  A Randomized Two-Stage Basket Trial Design to Optimize Doses for Multiple Indications}
\end{center}
\begin{center}
{\bf Shuqi Wang$^{1}$, Peter F. Thall$^{1}$, Kentaro Takeda$^{2}$, and Ying Yuan$^{1,*}$}
\end{center}

\begin{center}
$^{1}$Department of Biostatistics, The University of Texas MD Anderson Cancer Center\\
Houston, TX, USA.\\
$^{2}$Astellas Pharma Global Development Inc., Northbrook, IL, USA.\\
{*Email: yyuan@mdanderson.org}\\
\end{center}

\beginsupplement

\setcounter{equation}{0} \setcounter{table}{0} \setcounter{page}{1}
\setcounter{figure}{0}\setcounter{section}{0}


\section{ROMI design with more than two doses}

Denote the indications by $I_1,\cdots, I_K$ and index stages by $s=1,2.$ Suppose there are $J$ doses under investigation with $d_1 < \cdots < d_J$, where $d_J$ may be chosen as the MTD for a particular indication. In stage 1,  all patients are treated with $d_J$, and indications for which  $d_J$ is unsafe or ineffective are screened out. Indications that pass the stage 1 screening proceed to stage 2, where patients are randomized to doses $d_{\ell}$, $\ell=1,\dots, J$, and each dose is screened in each indication.  At the end of stage 2, for each indication with at least one acceptable dose, the OBD  is defined as the dose that maximizes the posterior mean utility. Stage 1 dose screening proceeds as described in Section 2.1 in the paper. In stage 2 dose optimization, the Bayesian hierarchical model needs to be revised when there are more than two doses. 

Define $Q_{\ell,k}$ as the standardized mean utilities and let $Z_{\ell,k}$ denote the number of quasi-events with quasi-probability $Q_{\ell,k}$ among $N_{\ell, k, 2}$ patients in $I_k$  treated with dose $d_{\ell}$ in stage 2, Detailed definition of $Q_{\ell,k}$ and $Z_{\ell,k}$ can be found in Section 2.2 of the paper, In $I_k$, We define the latent cluster variable $\zeta_k = g$ denoting that the $k$-th indication belongs to the $g$-th cluster where $d_{g}$ is the OBD for $g=1,\dots, J$, i.e., $\zeta_k = \arg\max_{\ell} Q_{\ell,k}$. Using ROMI version 1, where only stage 2 data is used for dose optimization, the following Bayesian hierarchical model can be applied: 
\begin{eqnarray} \label{eqn:BHM-supp-v1}
    \begin{aligned}
     &Z_{\ell, k} \mid Q_{\ell, k} \sim Quasi{\rm -}Binom(N_{\ell, k, 2}, Q_{\ell, k}),\ \ {\rm for}\ \  \ell = 1,\cdots, J, \ k=1,\cdots,K, \\
    &\theta_{\ell^*, k} = {\rm logit}(Q_{\ell^*,k})  -  {\rm logit}(Q_{J,k}), \ \ {\rm for}\ \  \ell^* = 1,\cdots, J-1,\\
    & \theta_{\ell^*, k} \mid \zeta_k = g \sim  N(\mu_g,  \tau^2), \ \ \ {\rm for}\ \ g = 1,\cdots,J, 
    \end{aligned}
\end{eqnarray}
where  ${\rm logit}(q) = {\rm log}\{q/(1-q)\}$ for $q\in[0,\ 1]$, and $\theta_{\ell^*, k}$ is the difference between log odds of the quasi-probabilities characterizing the $d_{\ell^*}$-versus-$d_J$ effect in $I_k$. Here, we use the highest dose as the reference but any dose can be used as reference, analogous to the baseline category logit model. We  assign the following priors to the model parameters: 
\begin{eqnarray} \label{eqn:general-BHM-v1-prior} \nonumber
    \begin{aligned}
           & \mu_g  \sim N(\tilde{\mu}_g, \tilde{\tau}_g^2),  \ {\rm for}\ \ g =0,1,  \ \ {\rm and}\ \  \tau^2  \sim IG(a,b),\\
    & Q_{J,k} \sim  Beta(c,d), \\
     &  \zeta_k \sim multinomial(q_1, \cdots, q_J),  \ \ {\rm and} \ \ q_1,\cdots,q_J \sim Dirichlet(\alpha_1, \cdots, \alpha_J),
    \end{aligned}
\end{eqnarray}

where $\tilde{\mu}_g, \tilde{\tau}_g^2, a, b, c, d, \alpha_1,\cdots, \alpha_K$ are fixed hyperparameters. Since   $\theta_{\ell^*, k}$ = ${\rm logit}(Q_{\ell^*,k})  -  {\rm logit}(Q_{J,k}),$
and normal priors are specified on the   $\theta_{\ell^*, k}$'s, the model is  completed by specifying the
priors on the $Q_{J,k}$'s.

For ROMI version 2,  we aim to use the data from both stages to determine the OBDs. To model the stage 1 data on $d_J$ and stage 2 data on $\{d_1, \cdots, d_J\}$, while accounting for the potential drift effects between stages, we can use the following extended Bayesian hierarchical model:
\begin{eqnarray} \label{eqn:BHM-supp-v2}
    \begin{aligned}
     &Z_{J, k, 1} \mid Q_{J, k, 1} \sim Quasi{\rm -}Binom(N_{J, k, 1}, Q_{J, k, 1}) \quad \quad\quad\quad\quad\quad\quad\quad\quad   (\rm \ {\bf Stage\ 1}\ )\ \\
     & Z_{\ell, k, 2} \mid Q_{\ell, k, 2} \sim Quasi{\rm -}Binom(N_{\ell, k, 2}, Q_{\ell, k, 2}), \ \ {\rm for}\ \ \ell = 1,\cdots, J, \ \ (\rm \ {\bf Stage\ 2}\ )\\
     &\beta_k = {\rm logit}(Q_{J, k, 1}) -  {\rm logit}(Q_{J, k, 2}),\\
     &\theta_{\ell^*, k,2}  = {\rm logit}(Q_{\ell^*,k,2})  -  {\rm logit}(Q_{H,k,2}), \ \ {\rm for}\ \  \ell^*=1,\cdots, J-1,\\
       & \theta_{\ell^*, k,2} \mid \zeta=g \sim iid\ N(\mu_g,  \tau^2), \ \ \ {\rm for}\ \ g = 1, \cdots, J,  
    \end{aligned}
\end{eqnarray}
where $\beta_k$ is defined as the $d_H$ between-stage drift parameter for each $I_k$, and $\theta_{\ell^*, k,2}$ denotes the between dose effects in stage 2. The following priors are assigned to the model parameters: 
\begin{eqnarray} \label{eqn:BHM-supp-v2-prior} \nonumber
    \begin{aligned}
    & \beta_k \sim\  \  \omega N(0,\sigma^2_{spike}) + (1-\omega)N(0,\sigma^2_{slab}), \\
    & \mu_g  \sim N(\tilde{\mu}_g, \tilde{\tau}_g^2),  \ {\rm for}\ \ g =0,1,  \ \ \tau^2  \sim IG(a,b), \quad {\rm and}\ \ \omega \sim U[0,\ 1],\\
        & Q_{J,k,2}  \sim  \ Beta(c,d),   \\
       &  \zeta_k \sim multinomial(q_1, \cdots, q_J),   \ \  {\rm and}\quad  q_1,\cdots,q_J \sim Dirichlet(\alpha_1, \cdots, \alpha_J).
    \end{aligned}
\end{eqnarray}

\section{Simulation results with three or six indications in the basket trial}
\begin{longtable}{llllllllll}
\caption{Simulation results of Pool, Independent, ROMI-v1-NC, ROMI-v1, and ROMI-v2 designs with three indications. Values for the true OBD of each indication are given in boldface. Doses are indexed by  $\ell = L, H$ and indications by $\ k=1,2,3.$ }
\label{tab:simulation-results-3_inds}\\
\hline
 &  & \multicolumn{6}{c}{Probability  (\%) of selecting the dose as OBD} &  &  \\ \cline{3-8}
 &  & \multicolumn{2}{c}{$I_1$} & \multicolumn{2}{c}{$I_2$} & \multicolumn{2}{c}{$I_3$} &  &  \\ \cline{3-8}
\multirow{-3}{*}{Design} & \multirow{-3}{*}{\textbf{}} & \multicolumn{1}{c}{$d_H$} & \multicolumn{1}{c}{$d_L$} & \multicolumn{1}{c}{$d_H$} & \multicolumn{1}{c}{$d_L$} & \multicolumn{1}{c}{$d_H$} & \multicolumn{1}{c}{$d_L$} & \multirow{-3}{*}{CSP} & \multirow{-3}{*}{N} \\ \hline
\endfirsthead
\multicolumn{10}{c}%
{ \thetable\ continued from previous page} \\
\hline
 &  & \multicolumn{6}{c}{Probability  (\%) of selecting the dose as OBD} &  &  \\ \cline{3-8}
 &  & \multicolumn{2}{c}{$I_1$} & \multicolumn{2}{c}{$I_2$} & \multicolumn{2}{c}{$I_3$} &  &  \\ \cline{3-8}
\multirow{-3}{*}{Design} & \multirow{-3}{*}{\textbf{}} & \multicolumn{1}{c}{$d_H$} & \multicolumn{1}{c}{$d_L$} & \multicolumn{1}{c}{$d_H$} & \multicolumn{1}{c}{$d_L$} & \multicolumn{1}{c}{$d_H$} & \multicolumn{1}{c}{$d_L$} & \multirow{-3}{*}{CSP} & \multirow{-3}{*}{N} \\ \hline
\endhead
\hline
\endfoot
\endlastfoot
\rowcolor[HTML]{EFEFEF}
\multicolumn{2}{l}{\cellcolor[HTML]{EFEFEF}Scenario A1} &  &  &  &  &  &  &  &  \\
\rowcolor[HTML]{EFEFEF}
 & $\pi_{T,\ell,k}^{true}$ & 0.4 & 0.3 & 0.4 & 0.3 & 0.4 & 0.3 &  &  \\
\rowcolor[HTML]{EFEFEF}
 & $\pi_{R,\ell,k}^{true}$ & 0.05 & 0.05 & 0.05 & 0.05 & 0.05 & 0.05 &  &  \\
\rowcolor[HTML]{EFEFEF}
 & $\bar{U}_{\ell,k}^{true}$ & 27 & 31 & 27 & 31 & 27 & 31 &  &  \\
\rowcolor[HTML]{EFEFEF}
Pool &  & 0.0 & 0.0 & 0.0 & 0.0 & 0.0 & 0.0 &NA  & 85 \\
\rowcolor[HTML]{EFEFEF}
Independent &  & 3.0 & 3.2 & 2.4 & 3.4 & 3.0 & 3.5 &NA  & 96 \\
\rowcolor[HTML]{EFEFEF}
ROMI-v1-NC &  & 0.4 & 1.0 & 0.6 & 0.7 & 0.6 & 0.9 &NA  & 54 \\
\rowcolor[HTML]{EFEFEF}
ROMI-v1 &  & 0.4 & 1.0 & 0.6 & 0.7 & 0.6 & 0.9 &NA  & 54 \\
\rowcolor[HTML]{EFEFEF}
ROMI-v2 &  & 0.4 & 1.0 & 0.6 & 0.8 & 0.6 & 0.9 &NA  & 54 \\
\multicolumn{2}{l}{Scenario A2} &  &  &  &  &  &  &  &  \\
 & $\pi_{T,\ell,k}^{true}$ & 0.25 & \textbf{0.15} & 0.25 & \textbf{0.15} & 0.25 & \textbf{0.15} &  &  \\
 & $\pi_{R,\ell,k}^{true}$ & 0.4 & \textbf{0.4} & 0.4 & \textbf{0.4} & 0.4 & \textbf{0.4} &  &  \\
 & $\bar{U}_{\ell,k}^{true}$& 54 & \textbf{58} & 54 & \textbf{58} & 54 & \textbf{58} &  &  \\
Pool &  & 20.4 & \textbf{79.7} & 20.4 & \textbf{79.7} & 20.4 & \textbf{79.7} & 79.7 & 162 \\
Independent &  & 32.1 & \textbf{68.0} & 30.5 & \textbf{69.5} & 32.8 & \textbf{67.2} & 68.2 & 162 \\
ROMI-NC &  & 23.8 & \textbf{75.1} & 24.4 & \textbf{74.8} & 23.5 & \textbf{75.3} & 75.1 & 161 \\
ROMI-v1 &  & 30.0 & \textbf{68.9} & 30.2 & \textbf{69.1} & 29.4 & \textbf{69.5} & 69.2 & 161 \\
ROMI-v2 &  & 27.9 & \textbf{71.0} & 27.1 & \textbf{72.1} & 27.8 & \textbf{71.1} & 71.4 & 161 \\
\rowcolor[HTML]{EFEFEF}
\multicolumn{2}{l}{Scenario A3} &  &  &  &  &  &  &  &  \\
\rowcolor[HTML]{EFEFEF}
 & $\pi_{T,\ell,k}^{true}$ & \textbf{0.25} & 0.15 & 0.4 & 0.3 & 0.4 & 0.3 &  &  \\
 \rowcolor[HTML]{EFEFEF}
 & $\pi_{R,\ell,k}^{true}$ & \textbf{0.4} & 0.3 & 0.05 & 0.05 & 0.05 & 0.05 &  &  \\
\rowcolor[HTML]{EFEFEF}
 & $\bar{U}_{\ell,k}^{true}$ & \textbf{56} & 52 & 27 & 31 & 27 & 31 &  &  \\
\rowcolor[HTML]{EFEFEF}
Pool &  & \textbf{28.2} & 7.3 & 28.2 & 7.3 & 28.2 & 7.3 & 28.2 & 119 \\
\rowcolor[HTML]{EFEFEF}
Independent &  & \textbf{70.6} & 29.4 & 2.2 & 2.7 & 2.5 & 2.8 & 70.6 & 117 \\
\rowcolor[HTML]{EFEFEF}
ROMI-v1-NC &  & \textbf{64.5} & 34.8 & 1.0 & 0.8 & 1.0 & 0.9 & 64.5 & 89 \\
\rowcolor[HTML]{EFEFEF}
ROMI-v1 &  & \textbf{64.6} & 34.8 & 1.0 & 0.8 & 1.0 & 0.9 & 64.6 & 89 \\
\rowcolor[HTML]{EFEFEF}
ROMI-v2 &  & \textbf{68.6} & 30.8 & 1.0 & 0.9 & 1.0 & 0.9 & 68.6 & 89 \\
\multicolumn{2}{l}{Scenario A4} &  &  &  &  &  &  &  &  \\
 & $\pi_{T,\ell,k}^{true}$ & 0.4 & 0.3 & \textbf{0.2} & 0.15 & \textbf{0.2} & 0.15 &  &  \\
 & $\pi_{R,\ell,k}^{true}$ & 0.05 & 0.05 & \textbf{0.4} & 0.3 & \textbf{0.4} & 0.3 &  &  \\
 & $\bar{U}_{\ell,k}^{true}$ & 27 & 31 & \textbf{56} & 52 & \textbf{56} & 52 &  &  \\
Pool &  & 64.3 & 35.2 & \textbf{64.3} & 35.2 & \textbf{64.3} & 35.2 & 64.3 & 156 \\
Independent &  & 2.8 & 3.2 & \textbf{68.2} & 31.9 & \textbf{69.3} & 30.7 & 68.7 & 139 \\
ROMI-v1-NC &  & 1.0 & 0.7 & \textbf{70.6} & 28.7 & \textbf{70.0} & 29.2 & 70.3 & 125 \\
ROMI-v1 &  & 1.0 & 0.7 & \textbf{67.7} & 31.6 & \textbf{67.9} & 31.2 & 67.8 & 125 \\
ROMI-v2 &  & 1.0 & 0.8 & \textbf{70.0} & 29.3 & \textbf{69.7} & 29.4 & 69.9 & 125 \\
\rowcolor[HTML]{F2F2F2} 
\multicolumn{2}{l}{\cellcolor[HTML]{EFEFEF}Scenario A5} &  &  &  &  &  &  &  &  \\
\rowcolor[HTML]{F2F2F2} 
\cellcolor[HTML]{EFEFEF} & \cellcolor[HTML]{EFEFEF}$\pi_{T,\ell,k}^{true}$ & 0.4 & 0.3 & \textbf{0.2} & 0.15 & 0.25 & \textbf{0.15} &  &  \\
\rowcolor[HTML]{F2F2F2} 
\cellcolor[HTML]{EFEFEF} & \cellcolor[HTML]{EFEFEF}$\pi_{R,\ell,k}^{true}$ & 0.05 & 0.05 & \textbf{0.4} & 0.3 & 0.4 & \textbf{0.4} &  &  \\
\rowcolor[HTML]{F2F2F2} 
\cellcolor[HTML]{EFEFEF} & \cellcolor[HTML]{EFEFEF}$\bar{U}_{\ell,k}^{true}$& 27 & 31 & \textbf{56} & 52 & 54 & \textbf{58} &  &  \\
\rowcolor[HTML]{F2F2F2} 
\cellcolor[HTML]{EFEFEF}Pool & \cellcolor[HTML]{EFEFEF} & 39.7 & 60.0 & \textbf{39.7} & 60.0 & 39.7 & \textbf{60.0} & 49.8 & 159 \\
\rowcolor[HTML]{F2F2F2} 
\cellcolor[HTML]{EFEFEF}Independent & \cellcolor[HTML]{EFEFEF} & 2.4 & 3.6 & \textbf{67.2} & 32.9 & 32.1 & \textbf{68.0} & 67.6 & 139 \\
\rowcolor[HTML]{F2F2F2} 
\cellcolor[HTML]{EFEFEF}ROMI-NC & \cellcolor[HTML]{EFEFEF} & 0.6 & 0.6 & \textbf{57.2} & 42.2 & 43.0 & \textbf{55.9} & 56.6 & 125 \\
\rowcolor[HTML]{F2F2F2} 
\cellcolor[HTML]{EFEFEF}ROMI-v1 & \cellcolor[HTML]{EFEFEF} & 0.6 & 0.6 & \textbf{62.8} & 36.4 & 36.0 & \textbf{62.9} & 62.9 & 125 \\
\rowcolor[HTML]{F2F2F2} 
\cellcolor[HTML]{EFEFEF}ROMI-v2 & \cellcolor[HTML]{EFEFEF} & 0.6 & 0.7 & \textbf{65.3} & 34.1 & 34.3 & \textbf{64.7} & 65.0 & 125 \\
Scenario A6 &  &  &  &  &  &  &  &  &  \\
 & $\pi_{T,\ell,k}^{true}$ & \textbf{0.2} & 0.15 & 0.25 & \textbf{0.15} & 0.25 & \textbf{0.15} &  &  \\
 & $\pi_{R,\ell,k}^{true}$ & \textbf{0.4} & 0.3 & 0.4 & \textbf{0.4} & 0.4 & \textbf{0.4} &  &  \\
 & $\bar{U}_{\ell,k}^{true}$ & \textbf{56} & 52 & 54 & \textbf{58} & 54 & \textbf{58} &  &  \\
Pool &  & \textbf{39.9} & 60.2 & 39.9 & \textbf{60.2} & 39.9 & \textbf{60.2} & 53.4 & 162 \\
Independent &  & \textbf{68.0} & 32.1 & 32.2 & \textbf{67.8} & 31.2 & \textbf{68.9} & 68.2 & 161 \\
ROMI-v1-NC &  & \textbf{47.4} & 52.1 & 36.5 & \textbf{62.6} & 35.8 & \textbf{63.0} & 57.7 & 161 \\
ROMI-v1 &  & \textbf{62.0} & 37.4 & 33.6 & \textbf{65.5} & 33.2 & \textbf{65.6} & 64.4 & 161 \\
ROMI-v2 &  & \textbf{62.4} & 37.1 & 32.2 & \textbf{67.0} & 32.8 & \textbf{66.1} & 65.2 & 161 \\ \hline
\end{longtable}
\vspace{-2em}
 \begin{tablenotes}
 \begin{centering}
      \small
      \item \hspace{0.5em} Abbreviations: CSP: correct selection percentage; N: average total sample size.
\end{centering}
    \end{tablenotes}
\newpage
\footnotesize
\newgeometry{left=1cm,right=1cm}
\begin{longtable}{llllllllllllllll}
\caption{Simulation results of Pool, Independent, ROMI-v1-NC, ROMI-v1, and ROMI-v2 designs with six indications. Values for the true OBD of each indication are given in boldface. Doses are indexed by  $\ell = L, H$ and indications by $\ k=1,2,3,4,5,6.$ }
\label{tab:simulation-results-6_inds}\\
\hline
 &  & \multicolumn{12}{c}{Probability  (\%) of selecting the dose as OBD} &  &  \\ \cline{3-14}
 &  & \multicolumn{2}{c}{$I_1$} & \multicolumn{2}{c}{$I_2$} & \multicolumn{2}{c}{$I_3$} & \multicolumn{2}{c}{$I_4$} & \multicolumn{2}{c}{$I_5$} & \multicolumn{2}{c}{$I_6$} &  &  \\ \cline{3-14}
\multirow{-3}{*}{Design} & \multirow{-3}{*}{\textbf{}} & \multicolumn{1}{c}{$d_H$} & \multicolumn{1}{c}{$d_L$} & \multicolumn{1}{c}{$d_H$} & \multicolumn{1}{c}{$d_L$} & \multicolumn{1}{c}{$d_H$} & \multicolumn{1}{c}{$d_L$} & \multicolumn{1}{c}{$d_H$} & \multicolumn{1}{c}{$d_L$} & \multicolumn{1}{c}{$d_H$} & \multicolumn{1}{c}{$d_L$} & \multicolumn{1}{c}{$d_H$} & \multicolumn{1}{c}{$d_L$} & \multirow{-3}{*}{CSP} & \multirow{-3}{*}{N} \\ \hline
\endfirsthead
\multicolumn{16}{c}%
{\thetable \ continued from previous page} \\
\hline
 &  & \multicolumn{12}{c}{Probability  (\%) of selecting the dose as OBD} &  &  \\ \cline{3-14}
 &  & \multicolumn{2}{c}{$I_1$} & \multicolumn{2}{c}{$I_2$} & \multicolumn{2}{c}{$I_3$} & \multicolumn{2}{c}{$I_4$} & \multicolumn{2}{c}{$I_5$} & \multicolumn{2}{c}{$I_6$} &  &  \\ \cline{3-14}
\multirow{-3}{*}{Design} & \multirow{-3}{*}{\textbf{}} & \multicolumn{1}{c}{$d_H$} & \multicolumn{1}{c}{$d_L$} & \multicolumn{1}{c}{$d_H$} & \multicolumn{1}{c}{$d_L$} & \multicolumn{1}{c}{$d_H$} & \multicolumn{1}{c}{$d_L$} & \multicolumn{1}{c}{$d_H$} & \multicolumn{1}{c}{$d_L$} & \multicolumn{1}{c}{$d_H$} & \multicolumn{1}{c}{$d_L$} & \multicolumn{1}{c}{$d_H$} & \multicolumn{1}{c}{$d_L$} & \multirow{-3}{*}{CSP} & \multirow{-3}{*}{N} \\ \hline
\endhead
\hline
\endfoot
\endlastfoot
\rowcolor[HTML]{EFEFEF} 
\multicolumn{2}{l}{\cellcolor[HTML]{EFEFEF}Scenario B1} &  &  &  &  &  &  &  &  &  &  &  &  &  &  \\
\rowcolor[HTML]{EFEFEF} 
 & $\pi_{T,\ell,k}^{true}$ & 0.4 & 0.3 & 0.4 & 0.3 & 0.4 & 0.3 & 0.4 & 0.3 & 0.4 & 0.3 & 0.4 & 0.3 &  &  \\
\rowcolor[HTML]{EFEFEF} 
 & $\pi_{R,\ell,k}^{true}$ & 0.05 & 0.05 & 0.05 & 0.05 & 0.05 & 0.05 & 0.05 & 0.05 & 0.05 & 0.05 & 0.05 & 0.05 &   &  \\
\rowcolor[HTML]{EFEFEF} 
 & $\bar{U}_{\ell,k}^{true}$ & 27 & 31 & 27 & 31 & 27 & 31 & 27 & 31 & 27 & 31 & 27 & 31 &   &  \\
\rowcolor[HTML]{EFEFEF} 
Pool &  & 0.0 & 0.0 & 0.0 & 0.0 & 0.0 & 0.0 & 0.0 & 0.0 & 0.0 & 0.0 & 0.0 & 0.0 &NA   & 168 \\
\rowcolor[HTML]{EFEFEF} 
Independent &  & 2.7 & 3.5 & 2.5 & 2.9 & 3.0 & 3.3 & 1.9 & 2.5 & 2.8 & 2.8 & 2.5 & 3.7 &NA   & 191 \\
\rowcolor[HTML]{EFEFEF} 
ROMI-v1-NC &  & 0.7 & 0.9 & 0.5 & 1.0 & 0.6 & 1.1 & 0.4 & 0.9 & 0.7 & 1.1 & 0.5 & 1.0 &NA   & 107 \\
\rowcolor[HTML]{EFEFEF} 
ROMI-v1 &  & 0.7 & 0.9 & 0.5 & 1.0 & 0.6 & 1.1 & 0.4 & 0.9 & 0.7 & 1.1 & 0.5 & 1.0 &NA   & 107 \\
\rowcolor[HTML]{EFEFEF} 
ROMI-v2 &  & 0.8 & 0.9 & 0.5 & 1.0 & 0.7 & 1.1 & 0.5 & 0.9 & 0.7 & 1.1 & 0.5 & 1.0 &NA   & 107 \\
\multicolumn{2}{l}{Scenario B2} &  &  &  &  &  &  &  &  &  &  & \textbf{} &  &  &  \\
 & $\pi_{T,\ell,k}^{true}$ & 0.25 & \textbf{0.15} & 0.25 & \textbf{0.15} & 0.25 & \textbf{0.15} & 0.25 & \textbf{0.15} & 0.25 & \textbf{0.15} & 0.25 & \textbf{0.15} &  &  \\
 & $\pi_{R,\ell,k}^{true}$ & 0.4 & \textbf{0.4} & 0.4 & \textbf{0.4} & 0.4 & \textbf{0.4} & 0.4 & \textbf{0.4} & 0.4 & \textbf{0.4} & 0.4 & \textbf{0.4} &  &  \\
 & $\bar{U}_{\ell,k}^{true}$ & 54 & \textbf{58} & 54 & \textbf{58} & 54 & \textbf{58} & 54 & \textbf{58} & 54 & \textbf{58} & 54 & \textbf{58} &  &  \\
Pool &  & 11.8 & \textbf{88.2} & 11.8 & \textbf{88.2} & 11.8 & \textbf{88.2} & 11.8 & \textbf{88.2} & 11.8 & \textbf{88.2} & 11.8 & \textbf{88.2} & 88.2 & 324 \\
Independent &  & 31 & \textbf{69.0} & 32.3 & \textbf{67.8} & 32.8 & \textbf{67.2} & 30.1 & \textbf{69.9} & 31.1 & \textbf{68.9} & 32.2 & \textbf{67.8} & 68.4 & 323 \\
ROMI-v1-NC &  & 15 & \textbf{83.9} & 15.6 & \textbf{83.7} & 15.4 & \textbf{83.5} & 15.1 & \textbf{84.0} & 14.4 & \textbf{84.9} & 15.3 & \textbf{83.9} & 84.0 & 322 \\
ROMI-v1 &  & 21.5 & \textbf{77.4} & 21.4 & \textbf{77.8} & 22.6 & \textbf{76.2} & 20.5 & \textbf{78.6} & 19.9 & \textbf{79.4} & 22.2 & \textbf{76.9} & 77.7 & 322 \\
ROMI-v2 &  & 18.85 & \textbf{80.1} & 17.8 & \textbf{81.4} & 17.9 & \textbf{81.0} & 18.6 & \textbf{80.6} & 17.0 & \textbf{82.4} & 18.6 & \textbf{80.7} & 81.0 & 322 \\
\rowcolor[HTML]{EFEFEF} 
\multicolumn{2}{l}{\cellcolor[HTML]{EFEFEF}Scenario B3} &  &  &  &  &  &  &  & \textbf{} &  & \textbf{} &  & \textbf{} &  &  \\
\rowcolor[HTML]{EFEFEF} 
 & $\pi_{T,\ell,k}^{true}$ & 0.4 & 0.3 & 0.4 & 0.3 & 0.4 & 0.3 & 0.4 & 0.3 & \textbf{0.2} & 0.15 & \textbf{0.2} & 0.15 &  &  \\
\rowcolor[HTML]{EFEFEF} 
 & $\pi_{R,\ell,k}^{true}$ & 0.05 & 0.05 & 0.05 & 0.05 & 0.05 & 0.05 & 0.05 & 0.05 & \textbf{0.4} & 0.3 & \textbf{0.4} & 0.3 &  &  \\
\rowcolor[HTML]{EFEFEF} 
 & $\bar{U}_{\ell,k}^{true}$ & 27 & 31 & 27 & 31 & 27 & 31 & 27 & 31 & \textbf{56} & 52 & \textbf{56} & 52 &  &  \\
\rowcolor[HTML]{EFEFEF} 
\cellcolor[HTML]{EFEFEF}Pool &  & 13.9 & 1.0 & 13.9 & 1.0 & 13.9 & 1.0 & 13.9 & 1.0 & \textbf{13.9} & 1.0 & \textbf{13.9} & 1.0 & 13.9 & 213 \\
\rowcolor[HTML]{EFEFEF} 
\cellcolor[HTML]{EFEFEF}Independent &  & 2.4 & 3.3 & 2.5 & 3.3 & 2.9 & 3.8 & 2.8 & 3.0 & \textbf{68.2} & 31.7 & \textbf{69.9} & 30.1 & 69.1 & 234 \\
\rowcolor[HTML]{EFEFEF} 
\cellcolor[HTML]{EFEFEF}ROMI-v1-NC &  & 0.7 & 1.1 & 1.0 & 0.8 & 0.7 & 1.2 & 0.9 & 1.0 & \textbf{68.8} & 30.4 & \textbf{68.7} & 30.5 & 68.8 & 178 \\
\rowcolor[HTML]{EFEFEF} 
\cellcolor[HTML]{EFEFEF}ROMI-v1 &  & 0.7 & 1.1 & 1.0 & 0.8 & 0.7 & 1.2 & 0.9 & 1.0 & \textbf{67.6} & 31.5 & \textbf{68.0} & 31.2 & 67.8 & 178 \\
\rowcolor[HTML]{EFEFEF} 
\cellcolor[HTML]{EFEFEF}ROMI-v2 &  & 0.8 & 1.1 & 1.0 & 0.8 & 0.7 & 1.2 & 0.9 & 1.0 & \textbf{69.3} & 29.9 & \textbf{69.2} & 30.1 & 69.2 & 178 \\
\multicolumn{2}{l}{Scenario B4} &  &  &  &  & \textbf{} &  & \textbf{} &  & \textbf{} &  & \textbf{} &  &  &  \\
 & $\pi_{T,\ell,k}^{true}$ & 0.4 & 0.3 & 0.4 & 0.3 & \textbf{0.2} & 0.15 & \textbf{0.2} & 0.15 & \textbf{0.2} & 0.15 & \textbf{0.2} & 0.15 &  &  \\
 & $\pi_{R,\ell,k}^{true}$ & 0.05 & 0.05 & 0.05 & 0.05 & \textbf{0.4} & 0.3 & \textbf{0.4} & 0.3 & \textbf{0.4} & 0.3 & \textbf{0.4} & 0.3 &  &  \\
 & $\bar{U}_{\ell,k}^{true}$ & 27 & 31 & 27 & 31 & \textbf{56} & 52 & \textbf{56} & 52 & \textbf{56} & 52 & \textbf{56} & 52 &  &  \\
Pool &  & 69.2 & 30.7 & 69.2 & 30.7 & \textbf{69.2} & 30.7 & \textbf{69.2} & 30.7 & \textbf{69.2} & 30.7 & \textbf{69.2} & 30.7 & 69.2 & 313 \\
Independent &  & 2.2 & 4.0 & 2.6 & 3.1 & \textbf{68.1} & 31.9 & \textbf{66.3} & 33.6 & \textbf{69.8} & 30.2 & \textbf{69.4} & 30.7 & 68.4 & 277 \\
ROMI-v1-NC &  & 0.9 & 0.9 & 0.6 & 1.0 & \textbf{76.6} & 22.6 & \textbf{76.9} & 22.2 & \textbf{76.0} & 23.2 & \textbf{77.4} & 21.9 & 76.7 & 249 \\
ROMI-v1 &  & 0.9 & 0.9 & 0.6 & 1.0 & \textbf{71.2} & 28.0 & \textbf{70.7} & 28.4 & \textbf{70.8} & 28.4 & \textbf{72.3} & 27.0 & 71.3 & 249 \\
ROMI-v2 &  & 0.9 & 0.9 & 0.6 & 1.0 & \textbf{74.4} & 24.8 & \textbf{75.2} & 24.0 & \textbf{73.7} & 25.6 & \textbf{75.9} & 23.5 & 74.8 & 249 \\
\rowcolor[HTML]{EFEFEF} 
\multicolumn{2}{l}{\cellcolor[HTML]{EFEFEF}Scenario B5} & \textbf{} &  & \textbf{} &  & \textbf{} &  & \textbf{} &  & \textbf{} &  & \textbf{} &  &  &  \\
\rowcolor[HTML]{EFEFEF} 
 & $\pi_{T,\ell,k}^{true}$ & 0.4 & 0.3 & 0.4 & 0.3 & 0.4 & 0.3 & 0.4 & 0.3 & \textbf{0.2} & 0.15 & 0.25 & \textbf{0.15} &  &  \\
\rowcolor[HTML]{EFEFEF} 
 & $\pi_{R,\ell,k}^{true}$ & 0.05 & 0.05 & 0.05 & 0.05 & 0.05 & 0.05 & 0.05 & 0.05 & \textbf{0.4} & 0.3 & 0.4 & \textbf{0.4} &  &  \\
\rowcolor[HTML]{EFEFEF} 
 & $\bar{U}_{\ell,k}^{true}$ & 27 & 31 & 27 & 31 & 27 & 31 & 27 & 31 & \textbf{56} & 52 & 54 & \textbf{58} &  &  \\
\rowcolor[HTML]{EFEFEF} 
\cellcolor[HTML]{EFEFEF}Pool &  & 11.9 & 4.1 & 11.9 & 4.1 & 11.9 & 4.1 & 11.9 & 4.1 & \textbf{11.9} & 4.1 & 11.9 & \textbf{4.1} & 8.0 & 223 \\
\rowcolor[HTML]{EFEFEF} 
\cellcolor[HTML]{EFEFEF}Independent &  & 2.9 & 3.3 & 2.5 & 3.7 & 3.0 & 2.9 & 2.4 & 2.9 & \textbf{68.2} & 31.7 & 32.0 & \textbf{68.0} & 68.1 & 234 \\
\rowcolor[HTML]{EFEFEF} 
\cellcolor[HTML]{EFEFEF}ROMI-v1-NC &  & 0.7 & 0.9 & 0.5 & 0.7 & 0.6 & 0.9 & 0.6 & 1.6 & \textbf{56.5} & 42.6 & 43.7 & \textbf{55.5} & 56.0 & 178 \\
\rowcolor[HTML]{EFEFEF} 
\cellcolor[HTML]{EFEFEF}ROMI-v1 &  & 0.6 & 1.0 & 0.5 & 0.7 & 0.6 & 0.9 & 0.6 & 1.6 & \textbf{63.9} & 35.2 & 36.7 & \textbf{62.5} & 63.2 & 178 \\
\rowcolor[HTML]{EFEFEF} 
\cellcolor[HTML]{EFEFEF}ROMI-v2 &  & 0.7 & 1.0 & 0.6 & 0.8 & 0.6 & 0.9 & 0.6 & 1.7 & \textbf{67.8} & 31.3 & 34.8 & \textbf{64.5} & 66.2 & 178 \\
Scenario B6 &  &  &  &  &  &  &  &  &  &  &  &  &  &  &  \\
 & $\pi_{T,\ell,k}^{true}$ & \textbf{0.2} & 0.15 & \textbf{0.2} & 0.15 & 0.25 & \textbf{0.15} & 0.25 & \textbf{0.15} & 0.25 & \textbf{0.15} & 0.25 & \textbf{0.15} &  &  \\
 & $\pi_{R,\ell,k}^{true}$ & \textbf{0.4} & 0.3 & \textbf{0.4} & 0.3 & 0.4 & \textbf{0.4} & 0.4 & \textbf{0.4} & 0.4 & \textbf{0.4} & 0.4 & \textbf{0.4} &  &  \\
 & $\bar{U}_{\ell,k}^{true}$ & \textbf{56} & 52 & \textbf{56} & 52 & 54 & \textbf{58} & 54 & \textbf{58} & 54 & \textbf{58} & 54 & \textbf{58} &  &  \\
Pool &  & \textbf{33.1} & 66.9 & \textbf{33.1} & 66.9 & 33.1 & \textbf{66.9} & 33.1 & \textbf{66.9} & 33.1 & \textbf{66.9} & 33.1 & \textbf{66.9} & 55.6 & 324 \\
Independent &  & \textbf{69.7} & 30.3 & \textbf{69.1} & 30.9 & 32.0 & \textbf{68.1} & 30.7 & \textbf{69.3} & 33.7 & \textbf{66.4} & 32.4 & \textbf{67.7} & 68.4 & 322 \\
ROMI-v1-NC &  & \textbf{41.6} & 57.9 & \textbf{41.7} & 57.6 & 32.1 & \textbf{66.8} & 32.6 & \textbf{66.4} & 32.2 & \textbf{67.1} & 32.3 & \textbf{66.8} & 58.4 & 321 \\
ROMI-v1 &  & \textbf{55.5} & 44.0 & \textbf{55.1} & 44.1 & 31.9 & \textbf{67.0} & 31.8 & \textbf{67.3} & 29.6 & \textbf{69.7} & 30.3 & \textbf{68.9} & 63.9 & 321 \\
ROMI-v2 &  & \textbf{56.0} & 43.5 & \textbf{53.8} & 45.5 & 30.1 & \textbf{68.9} & 29.6 & \textbf{69.5} & 29.1 & \textbf{70.2} & 30.0 & \textbf{69.2} & 64.6 & 321 \\ \hline
\end{longtable}
\vspace{-1.5em}
 \begin{tablenotes}
 \begin{centering}
      \small
      \item Abbreviations: CSP: correct selection percentage; N: average total sample size.
\end{centering}
    \end{tablenotes}

\newpage
\restoregeometry
\section{ROMI design when prior for the shrinkage parameter follows Half-Cauchy distribution}
\begin{longtable}[c]{llcccccccccc}
\caption{Simulation results for ROMI-v1 and ROMI-v2 designs when the prior assigned to the shrinkage parameter $\tau^2$ is Half-Cauchy distribution  with scale 25. Values for the true OBD of each indication are given in boldface. Doses are indexed by  $\ell = L, H$ and indications by $\ k=1,2,3,4.$ 
}
\label{tab:simulation-results-halfcauchy}\\
\hline
 &  & \multicolumn{8}{l}{Probability  (\%) of selecting the dose as OBD} & \multicolumn{1}{l}{} & \multicolumn{1}{l}{} \\ \cline{3-10}
 &  & \multicolumn{2}{c}{$I_1$} & \multicolumn{2}{c}{$I_2$} & \multicolumn{2}{c}{$I_3$} & \multicolumn{2}{c}{$I_4$} & \multicolumn{1}{l}{} & \multicolumn{1}{l}{} \\ \cline{3-10}
\multirow{-3}{*}{Design} & \multirow{-3}{*}{\textbf{}} & $d_H$ & $d_L$  & $d_H$  & $d_L$ & $d_H$  & $d_L$ & $d_H$  & $d_L$ & \multicolumn{1}{l}{\multirow{-3}{*}{CSP}} & \multicolumn{1}{l}{\multirow{-3}{*}{N}} \\ \hline
\endfirsthead
\multicolumn{12}{c}%
{ \thetable\ continued from previous page} \\
\hline
 &  & \multicolumn{8}{l}{Probability (\%) of selecting the dose as OBD} & \multicolumn{1}{l}{} & \multicolumn{1}{l}{} \\ \cline{3-10}
 &  & \multicolumn{2}{c}{$I_1$} & \multicolumn{2}{c}{$I_2$} & \multicolumn{2}{c}{$I_3$} & \multicolumn{2}{c}{$I_4$} & \multicolumn{1}{l}{} & \multicolumn{1}{l}{} \\ \cline{3-10}
\multirow{-3}{*}{Design} & \multirow{-3}{*}{\textbf{}} & $d_H$  & $d_L$ & $d_H$  & $d_L$ & $d_H$  & $d_L$ & $d_H$  & $d_L$ & \multicolumn{1}{l}{\multirow{-3}{*}{CSP}} & \multicolumn{1}{l}{\multirow{-3}{*}{N}} \\ \hline
\endhead
\hline
\endfoot
\endlastfoot
\rowcolor[HTML]{EFEFEF} 
Scenario 1 &  &  &  &  &  &  &  &  &  &  &  \\
\rowcolor[HTML]{EFEFEF} 
 & $\pi_{T,\ell,k}^{true}$ & 0.4 & 0.3 & 0.4 & 0.3 & 0.4 & 0.3 & 0.4 & 0.3 &  &  \\
\rowcolor[HTML]{EFEFEF} 
 & $\pi_{R,\ell,k}^{true}$& 0.05 & 0.05 & 0.05 & 0.05 & 0.05 & 0.05 & 0.05 & 0.05 &  &  \\
\rowcolor[HTML]{EFEFEF} 
 & $\bar{U}_{\ell,k}^{true}$ & 27 & 31 & 27 & 31 & 27 & 31 & 27 & 31 &  &  \\
\rowcolor[HTML]{EFEFEF} 
ROMI-v1 &  & 1.3 & 0.9 & 0.5 & 0.6 & 0.6 & 0.7 & 1.3 & 0.7 &NA   & 71 \\
\rowcolor[HTML]{EFEFEF} 
ROMI-v2 &  & 1.3 & 0.9 & 0.5 & 0.7 & 0.7 & 0.8 & 1.3 & 0.8 &NA   & 71 \\
Scenario 2 &  &  &  &  &  &  &  &  &  &  &  \\
 & $\pi_{T,\ell,k}^{true}$ & \textbf{0.2} & 0.15 & 0.4 & 0.3 & 0.4 & 0.3 & 0.4 & 0.3 &  &  \\
 & $\pi_{R,\ell,k}^{true}$& \textbf{0.4} & 0.3 & 0.05 & 0.05 & 0.05 & 0.05 & 0.05 & 0.05 &  &  \\
 & $\bar{U}_{\ell,k}^{true}$ & \textbf{56} & 52 & 27 & 31 & 27 & 31 & 27 & 31 &  &  \\
ROMI-v1 &  & \textbf{65.1} & 34.4 & 0.7 & 1.0 & 0.9 & 0.8 & 1.0 & 0.7 & 65.1 & 107 \\
ROMI-v2 &  & \textbf{69.9} & 29.6 & 0.7 & 1.1 & 0.9 & 0.8 & 1.0 & 0.8 & 69.9 & 107 \\
\rowcolor[HTML]{EFEFEF} 
Scenario 3 &  &  &  &  &  &  &  &  &  &  &  \\
\rowcolor[HTML]{EFEFEF} 
 & $\pi_{T,\ell,k}^{true}$ & 0.25 & \textbf{0.15} & 0.4 & 0.3 & 0.4 & 0.3 & 0.4 & 0.3 &  &  \\
\rowcolor[HTML]{EFEFEF} 
 & $\pi_{R,\ell,k}^{true}$& 0.4 & \textbf{0.4} & 0.05 & 0.05 & 0.05 & 0.05 & 0.05 & 0.05 &  &  \\
\rowcolor[HTML]{EFEFEF} 
 & $\bar{U}_{\ell,k}^{true}$ & 54 & \textbf{58} & 27 & 31 & 27 & 31 & 27 & 31 &  &  \\
\rowcolor[HTML]{EFEFEF} 
\cellcolor[HTML]{EFEFEF}ROMI-v1 &  & 30.6 & \textbf{68.3} & 1.0 & 0.5 & 0.9 & 0.6 & 1.2 & 1.0 & 68.3 & 107 \\
\rowcolor[HTML]{EFEFEF} 
\cellcolor[HTML]{EFEFEF}ROMI-v2 &  & 30.0 & \textbf{68.9} & 0.9 & 0.6 & 0.9 & 0.7 & 1.2 & 1.0 & 68.9 & 107 \\
Scenario 4 &  & \multicolumn{1}{l}{} & \multicolumn{1}{l}{} & \multicolumn{1}{l}{} & \multicolumn{1}{l}{} & \multicolumn{1}{l}{} & \multicolumn{1}{l}{} & \multicolumn{1}{l}{} & \multicolumn{1}{l}{} & \multicolumn{1}{l}{} & \multicolumn{1}{l}{} \\
 & $\pi_{T,\ell,k}^{true}$ & \textbf{0.2} & 0.15 & 0.4 & 0.3 & 0.4 & 0.3 & \textbf{0.2} & 0.15 & \multicolumn{1}{l}{} & \multicolumn{1}{l}{} \\
 & $\pi_{R,\ell,k}^{true}$& \textbf{0.4} & 0.3 & 0.05 & 0.05 & 0.05 & 0.05 & \textbf{0.4} & 0.3 & \multicolumn{1}{l}{} & \multicolumn{1}{l}{} \\
 & $\bar{U}_{\ell,k}^{true}$ & \textbf{56} & 52 & 27 & 31 & 27 & 31 & \textbf{56} & 52 & \multicolumn{1}{l}{} & \multicolumn{1}{l}{} \\
ROMI-v1 &  & \textbf{68.5} & 31.0 & 0.7 & 0.8 & 1.0 & 1.1 & \textbf{64.1} & 35.1 & \multicolumn{1}{l}{66.3} & \multicolumn{1}{l}{143} \\
ROMI-v2 &  & \textbf{71.5} & 28.0 & 0.7 & 0.9 & 1.0 & 1.1 & \textbf{67.2} & 32.1 & \multicolumn{1}{l}{69.3} & \multicolumn{1}{l}{143} \\
\rowcolor[HTML]{EFEFEF} 
Scenario 5 &  &  &  &  &  &  &  &  &  & \multicolumn{1}{l}{\cellcolor[HTML]{EFEFEF}} & \multicolumn{1}{l}{\cellcolor[HTML]{EFEFEF}} \\
\rowcolor[HTML]{EFEFEF} 
 & $\pi_{T,\ell,k}^{true}$ & 0.25 & \textbf{0.15} & 0.4 & 0.3 & 0.4 & 0.3 & 0.25 & \textbf{0.15} & \multicolumn{1}{l}{\cellcolor[HTML]{EFEFEF}} & \multicolumn{1}{l}{\cellcolor[HTML]{EFEFEF}} \\
\rowcolor[HTML]{EFEFEF} 
 & $\pi_{R,\ell,k}^{true}$& 0.4 & \textbf{0.4} & 0.05 & 0.05 & 0.05 & 0.05 & 0.4 & \textbf{0.4} & \multicolumn{1}{l}{\cellcolor[HTML]{EFEFEF}} & \multicolumn{1}{l}{\cellcolor[HTML]{EFEFEF}} \\
\rowcolor[HTML]{EFEFEF} 
 & $\bar{U}_{\ell,k}^{true}$ & 54 & \textbf{58} & 27 & 31 & 27 & 31 & 54 & \textbf{58} &  &  \\
\rowcolor[HTML]{EFEFEF} 
\cellcolor[HTML]{EFEFEF}ROMI-v1 &  & 31.0 & \textbf{67.9} & 0.5 & 1.2 & 1.1 & 0.9 & 29.0 & \textbf{70.1} & 69.0 & 143 \\
\rowcolor[HTML]{EFEFEF} 
\cellcolor[HTML]{EFEFEF}ROMI-v2 &  & 32.7 & \textbf{66.3} & 0.6 & 1.2 & 1.1 & 0.9 & 29.9 & \textbf{69.2} & 67.7 & 143 \\
Scenario 6 &  &  &  &  &  &  &  &  &  &  &  \\
 & $\pi_{T,\ell,k}^{true}$ & 0.4 & 0.3 & \textbf{0.2} & 0.15 & \textbf{0.2} & 0.15 & \textbf{0.2} & 0.15 &  &  \\
 & $\pi_{R,\ell,k}^{true}$& 0.05 & 0.05 & \textbf{0.4} & 0.3 & \textbf{0.4} & 0.3 & \textbf{0.4} & 0.3 &  &  \\
 & $\bar{U}_{\ell,k}^{true}$ & 27 & 31 & \textbf{56} & 52 & \textbf{56} & 52 & \textbf{56} & 52 &  &  \\
ROMI-v1 &  & 1.2 & 1.0 & \textbf{68.3} & 31.0 & \textbf{66.5} & 32.5 & \textbf{67.2} & 31.9 & 67.3 & 178 \\
ROMI-v2 &  & 1.2 & 1.0 & \textbf{68.7} & 30.7 & \textbf{67.3} & 31.8 & \textbf{68.3} & 30.9 & 68.1 & 178 \\
\rowcolor[HTML]{EFEFEF} 
Scenario 7 &  & \multicolumn{1}{l}{\cellcolor[HTML]{EFEFEF}} & \multicolumn{1}{l}{\cellcolor[HTML]{EFEFEF}} & \multicolumn{1}{l}{\cellcolor[HTML]{EFEFEF}\textbf{}} & \multicolumn{1}{l}{\cellcolor[HTML]{EFEFEF}} & \multicolumn{1}{l}{\cellcolor[HTML]{EFEFEF}\textbf{}} & \multicolumn{1}{l}{\cellcolor[HTML]{EFEFEF}} & \multicolumn{1}{l}{\cellcolor[HTML]{EFEFEF}\textbf{}} & \multicolumn{1}{l}{\cellcolor[HTML]{EFEFEF}} & \multicolumn{1}{l}{\cellcolor[HTML]{EFEFEF}} & \multicolumn{1}{l}{\cellcolor[HTML]{EFEFEF}} \\
\rowcolor[HTML]{EFEFEF} 
 & $\pi_{T,\ell,k}^{true}$ & 0.4 & 0.3 & 0.25 & \textbf{0.15} & 0.25 & \textbf{0.15} & 0.25 & \textbf{0.15} &  &  \\
\rowcolor[HTML]{EFEFEF} 
 & $\pi_{R,\ell,k}^{true}$& 0.05 & 0.05 & 0.4 & \textbf{0.4} & 0.4 & \textbf{0.4} & 0.4 & \textbf{0.4} &  &  \\
\rowcolor[HTML]{EFEFEF} 
 & $\bar{U}_{\ell,k}^{true}$ & 27 & 31 & 54 & \textbf{58} & 54 & \textbf{58} & 54 & \textbf{58} &  &  \\
\rowcolor[HTML]{EFEFEF} 
\cellcolor[HTML]{EFEFEF}ROMI-v1 &  & 1.3 & 1.0 & 30.2 & \textbf{69.0} & 29.4 & \textbf{69.4} & 28.3 & \textbf{70.9} & 69.8 & 179 \\
\rowcolor[HTML]{EFEFEF} 
\cellcolor[HTML]{EFEFEF}ROMI-v2 &  & 1.3 & 0.9 & 32.5 & \textbf{66.8} & 32.5 & \textbf{66.4} & 30.5 & \textbf{68.7} & 67.3 & 179 \\
Scenario 8 &  &  &  &  & \textbf{} &  & \textbf{} &  & \textbf{} &  &  \\
 & $\pi_{T,\ell,k}^{true}$ & 0.4 & 0.3 & \textbf{0.2} & 0.15 & 0.25 & \textbf{0.15} & 0.25 & \textbf{0.15} &  &  \\
 & $\pi_{R,\ell,k}^{true}$& 0.05 & 0.05 & \textbf{0.4} & 0.3 & 0.4 & \textbf{0.4} & 0.4 & \textbf{0.4} &  & \textbf{} \\
 & $\bar{U}_{\ell,k}^{true}$ & 27 & 31 & \textbf{56} & 52 & 54 & \textbf{58} & 54 & \textbf{58} &  &  \\
ROMI-v1 &  & 1.4 & 1.0 & \textbf{61.6} & 37.8 & 32.8 & \textbf{66.1} & 32.2 & \textbf{67.0} & 64.9 & 179 \\
ROMI-v2 &  & 1.4 & 1.0 & \textbf{68.3} & 31.1 & 32.5 & \textbf{66.4} & 30.9 & \textbf{68.3} & 67.6 & 179 \\
\rowcolor[HTML]{EFEFEF} 
Scenario 9 &  &  &  & \textbf{} &  &  & \textbf{} &  & \textbf{} &  &  \\
\rowcolor[HTML]{EFEFEF} 
 & $\pi_{T,\ell,k}^{true}$ & \textbf{0.2} & 0.15 & \textbf{0.2} & 0.15 & \textbf{0.2} & 0.15 & \textbf{0.2} & 0.15 &  &  \\
\rowcolor[HTML]{EFEFEF} 
 & $\pi_{R,\ell,k}^{true}$& \textbf{0.4} & 0.3 & \textbf{0.4} & 0.3 & \textbf{0.4} & 0.3 & \textbf{0.4} & 0.3 &  &  \\
\rowcolor[HTML]{EFEFEF} 
 & $\bar{U}_{\ell,k}^{true}$ & \textbf{56} & 52 & \textbf{56} & 52 & \textbf{56} & 52 & \textbf{56} & 52 &  &  \\
\rowcolor[HTML]{EFEFEF} 
\cellcolor[HTML]{EFEFEF}ROMI-v1 &  & \textbf{69.7} & 29.8 & \textbf{68.9} & 30.6 & \textbf{68.9} & 30.2 & \textbf{68.6} & 30.5 & 69.0 & 214 \\
\rowcolor[HTML]{EFEFEF} 
\cellcolor[HTML]{EFEFEF}ROMI-v2 &  & \textbf{69.9} & 29.6 & \textbf{68.3} & 31.2 & \textbf{69.5} & 29.7 & \textbf{68.3} & 30.9 & 69.0 & 214 \\
Scenario 10 &  & \multicolumn{1}{l}{\textbf{}} & \multicolumn{1}{l}{} & \multicolumn{1}{l}{\textbf{}} & \multicolumn{1}{l}{} & \multicolumn{1}{l}{\textbf{}} & \multicolumn{1}{l}{} & \multicolumn{1}{l}{\textbf{}} & \multicolumn{1}{l}{} & \multicolumn{1}{l}{} & \multicolumn{1}{l}{} \\
 & $\pi_{T,\ell,k}^{true}$ & 0.25 & \textbf{0.15} & 0.25 & \textbf{0.15} & 0.25 & \textbf{0.15} & 0.25 & \textbf{0.15} &  &  \\
 & $\pi_{R,\ell,k}^{true}$& 0.4 & \textbf{0.4} & 0.4 & \textbf{0.4} & 0.4 & \textbf{0.4} & 0.4 & \textbf{0.4} &  &  \\
 & $\bar{U}_{\ell,k}^{true}$ & 54 & \textbf{58} & 54 & \textbf{58} & 54 & \textbf{58} & 54 & \textbf{58} &  &  \\
ROMI-v1 &  & 28.9 & \textbf{70.0} & 28.1 & \textbf{71.1} & 30.6 & \textbf{68.2} & 28.4 & \textbf{70.7} & 70.0 & 214 \\
ROMI-v2 &  & 31.9 & \textbf{67.0} & 30.7 & \textbf{68.5} & 32.1 & \textbf{66.8} & 31.1 & \textbf{68.1} & 67.6 & 214 \\
\rowcolor[HTML]{EFEFEF} 
Scenario 11 &  & \textbf{} &  &  & \textbf{} &  & \textbf{} &  & \textbf{} &  &  \\
\rowcolor[HTML]{EFEFEF} 
 & $\pi_{T,\ell,k}^{true}$ & \textbf{0.2} & 0.15 & \textbf{0.2} & 0.15 & 0.25 & \textbf{0.15} & 0.25 & \textbf{0.15} &  &  \\
\rowcolor[HTML]{EFEFEF} 
 & $\pi_{R,\ell,k}^{true}$& \textbf{0.4} & 0.3 & \textbf{0.4} & 0.3 & 0.4 & \textbf{0.4} & 0.4 & \textbf{0.4} &  &  \\
\rowcolor[HTML]{EFEFEF} 
 & $\bar{U}_{\ell,k}^{true}$ & \textbf{56} & 52 & \textbf{56} & 52 & 54 & \textbf{58} & 54 & \textbf{58} &  &  \\
\rowcolor[HTML]{EFEFEF} 
ROMI-v1 &  & \textbf{64.1} & 35.2 & \textbf{63.5} & 35.9 & 34.2 & \textbf{64.7} & 34.4 & \textbf{64.6} & 64.2 & 214 \\
\rowcolor[HTML]{EFEFEF} 
ROMI-v2 &  & \textbf{68.0} & 31.4 & \textbf{67.9} & 31.5 & 32.3 & \textbf{66.7} & 33.1 & \textbf{66.1} & 67.2 & 214 \\ \hline
\end{longtable}
\vspace{-1.5em}
 \begin{tablenotes}
 \begin{centering}
      \small
      \item Abbreviations: CSP: correct selection percentage; N: average total sample size.
\end{centering}
    \end{tablenotes}

\end{document}